\documentclass[11pt]{article}
\usepackage{amssymb,amsmath,amsthm,amscd,bbm}   % AMS packages
\usepackage{graphicx}  
\usepackage{here}                                               
\usepackage{psfrag}

\catcode`@=11
\renewcommand{\section}{\@startsection{section}{1}{0pt}{\medskipamount}
{\medskipamount}{\large\bf}}
\renewcommand{\subsection}{\@startsection{subsection}{1}{0pt}{\smallskipamount}
{\smallskipamount}{\normalsize\bf}}
\numberwithin{equation}{section}
\catcode`@=12

\newcommand{\mbf}[1]{{\boldsymbol{#1}}}
\newcommand{\unity}{\mathbbm{1}}
\newcommand{\R}{{\mathbbm{R}}}
\newcommand{\C}{{\mathbbm{C}}}
\newcommand{\Z}{{\mathbbm{Z}}}
\newcommand{\NN}{{\mathbbm{N}}}
\newcommand{\Hcal}{{\mathcal H}}
\newcommand{\Dcal}{{\mathcal D}}
\newcommand{\Ecal}{{\mathcal E}}
\newcommand{\adag}{a^{\dagger}}

\def\a{\alpha}
\def\ab{{\bar\alpha}}
\def\de{\delta}
\def\th{\theta}
\def\>{\rangle}
\def\<{\langle}
\def\={\ =\ }
\def\+{\dagger}
\def\pa{\partial}
\def\diff{\mathrm{d}}
\def\e{\mathrm{e}}
\def\ic{\mathrm{i}}
\def\tr{\mathrm{tr}}
\def\Tr{\mathrm{Tr}}
\def\sfrac#1#2{{\textstyle\frac{#1}{#2}}}

\begin{document}
\begin{titlepage}
\setcounter{page}{0}
\begin{flushright}
      hep-th/0412001\\
      ITP--UH-25/04\\
\end{flushright}
\vskip 2.0cm

\begin{center}
{\LARGE\bf  
Sigma-Model Solitons in the Noncommutative Plane: \\[8pt]
Construction and Stability Analysis 
}
\\
\vspace{14mm}
{\Large
Andrei V. Domrin}
\\[5mm]
{ \em
Department of Mathematics and Mechanics, Moscow State University\\
Leninskie gory, 119992, GSP-2, Moscow, Russia}
\\[5mm]
email: \texttt{domrin@mi.ras.ru}
\\[12mm]
{\Large
Olaf Lechtenfeld} \ \ and \ \
{\Large
Stefan Petersen}
\\[5mm]
{ \em
Institut f\"ur Theoretische Physik, Universit\"at Hannover \\
Appelstra\ss{}e 2, 30167 Hannover, Germany }
\\[5mm]
email: \texttt{lechtenf, petersen @itp.uni-hannover.de}
\\[12mm]
%\small \today
\end{center}
\vspace{15mm}

\begin{abstract}
\noindent
Noncommutative multi-solitons are investigated in Euclidean two-dimensional
U($n$) and Grassmannian sigma models, using the auxiliary Fock-space formalism.
Their construction and moduli spaces are reviewed in some detail, 
unifying abelian and nonabelian configurations.
The analysis of linear perturbations around these backgrounds reveals 
an unstable mode for the U($n$) models but shows stability 
for the Grassmannian case. For multi-solitons which are diagonal
in the Fock-space basis we explicitly evaluate the spectrum of the Hessian
and identify all zero modes. It is very suggestive but remains to be proven 
that our results qualitatively extend to the entire multi-soliton moduli space.
\end{abstract}

\vfill
\end{titlepage}

\tableofcontents

\pagebreak

\section{Introduction } 
\noindent
Multi-solitons in noncommutative Euclidean two-dimensional sigma models are 
of interest as static D0-branes inside D2-branes with a constant B-field 
background~\cite{Lechtenfeld:2001uq} but also per se as nonperturbative 
classical field configurations. Adding a temporal dimension, they represent 
static solutions not only of a (WZW-extended) sigma model but also of the 
Yang-Mills-Higgs BPS equations on noncommutative $\R^{2+1}$. In fact, 
the full BPS sector of the Yang-Mills-Higgs system is, in a particular gauge, 
given by the (time-dependent) solutions to the sigma-model equations of motion.

In the commutative case, the classical solutions of Euclidean two-dimensional
sigma models have been investigated intensively by physicists as well as
by mathematicians (for a review see, e.g.~\cite{Zakrzewski}).
Prominent target spaces are U($n$) group manifolds or their Grassmannian
cosets Gr$(n,r)=\frac{\textrm{U}(n)}{\textrm{U}(r){\times}\textrm{U}(n{-}r)}$
for $1\le r<n$ (which are geodesic submanifolds of U($n$)).\footnote{
Most familiar are the Gr$(n,1)=\C P^{n-1}$ models.} 
Any classical solution~$\Phi$ for the U($n$) sigma model can be constructed
iteratively, with at most $n{-}1$ so-called unitons as building blocks
\cite{Uhlenbeck, Wood}.
The subset of hermitian solutions, $\Phi^\+=\Phi$, coincides with the space
of Grassmannian solutions, $\Phi\in\text{Gr}(n,r)$ for some rank $r<n$, 
of which again a subset is distinguished by a BPS property.\footnote{
not to be confused with the BPS condition for the Yang-Mills-Higgs system}
These BPS configurations are precisely the one-uniton solutions and can be
interpreted as (static) multi-solitons.

Let us now perturb such multi-solitons within the configuration space
of the two-dimensional (i.e.~static) sigma-model, either within their
Grassmannian or, more widely, within the whole group manifold. 
A linear stability analysis then admits a two-fold interpretation. 
First, it is relevant for the semiclassical evaluation of
the Euclidean path integral, revealing potential {\it quantum\/} instabilities
of the two-dimensional model. Second, it yields the (infinitesimal) time
evolution of fluctuations around the static multi-soliton in the time-extended
three-dimensional theory, indicating {\it classical\/} instabilities if they
are present. More concretely, any static perturbation of a classical 
configuration can be taken as (part of the) Cauchy data for a classical time
evolution, and any negative eigenvalue of the quadratic fluctuation operator
will give rise to an exponential runaway behavior, at least within the linear
response regime. Furthermore, fluctuation zero modes are expected to belong
to moduli perturbations of the classical configuration under consideration.
The current knowledge on the effect of quantum fluctuations is summarized in
\cite{Zakrzewski}.

The Moyal deformation of Euclidean two-dimensional sigma models has been
described in~\cite{Lechtenfeld:2001aw}, 
including the construction of their BPS solutions.
The operator formulation of the noncommutative U($n$) theory turns it into a
(zero-dimensional) matrix model with U($\C^n{\otimes}\Hcal$) as its target 
space, where $\Hcal$ denotes a Heisenberg algebra representation module.
Each BPS configuration again belongs to some Grassmannian, whose rank~$r$ 
can be finite or infinite. The latter case represents a smooth
deformation of the known commutative multi-solitons, while the former situation
realizes a noncommutative novelty, namely abelian multi-solitons
(they even occur for the U(1) model). 
Although these solutions are available in explicit form, very little is known
about their stability.\footnote{ A few recent works \cite{Solovyov:2000xy,
Aganagic:2000mh,Gross:2000ss,Hadasz:2001cn,Fujii:2001wp,Durhuus:2001nj} 
address perturbations of noncommutative solitons but not for sigma models.}
Our work sheds some light on this issue, 
in particular for the interesting abelian case.

The paper is organized as follows. In the next section we review the
noncommutative two-dimensional U($n$) sigma-model and present nested 
classes of finite-energy solutions to its equation of motion. We focus
on BPS configurations (multi-solitons) and their moduli spaces, unifying 
the previously different descriptions of abelian and nonabelian multi-solitons.
Section~3 is devoted to a study of the fluctuations around generic 
multi-solitons and detects a universal unstable mode.
Section~4 analyzes the spectrum of the Hessian specifically for 
U(1)~backgrounds and settles the issue for diagonal BPS solutions,
analytically as well as numerically.
Section~5 addresses the fluctuation problem in the nonabelian case
for the example of the noncommutative U(2)~model.
We close with a summary and a list of open questions.

\section{Noncommutative 2d sigma model and its solutions}
\noindent
Before addressing fluctuations, it is necessary to present the 
noncommutative two-dimensional sigma model and its multi-soliton
solutions which we will set out to perturb later on.
These solutions have been discussed as static solutions of 
a $2{+}1$ dimensional noncommutative sigma model
\cite{Lechtenfeld:2001aw,Lechtenfeld:2001gf,Wolf:2002jw,Ihl:2002kz} 
which results from Moyal deforming the WZW-modified integrable sigma model
\cite{Ward:1988ie,Ward:1990vc,Ioannidou:1998jh}. 
Here we review the results pertaining to the static situation.

\subsection{Noncommutativity}
\noindent
A Moyal deformation of Euclidean $\R^2$ with coordinates $(x,y)$
is achieved by replacing the ordinary pointwise product of smooth functions 
on it with the noncommutative but associative Moyal star product. 
The latter is characterized by a constant positive real parameter~$\th$ which 
prominently appears in the star commutation relation between the coordinates,
\begin{equation} \label{nccoord}
x\star y - y\star x \ \equiv\ [\,x\,,\,y\,]_\star \= \ic\,\th \ .
\end{equation}
For a concise treatment of the Moyal star product 
see~\cite{Harvey:2001yn,Douglas:2001ba,Szabo:2001kg}.
It is convenient to work with the complex combinations
\begin{equation}
z \= x+\ic y \quad\textrm{and}\quad \bar{z} \= x-\ic y 
\qquad\Longrightarrow\qquad [\,z\,,\,\bar{z}\,]_\star \= 2\th 
\end{equation}
and to scale them to
\begin{equation}
a \= \tfrac{z}{\sqrt{2\th}} \quad\textrm{and}\quad 
a^\dagger \= \tfrac{\bar{z}}{\sqrt{2\th}}
\qquad\Longrightarrow\qquad [\,a\,,\,a^\dagger\,]_\star \= 1 \ . 
\end{equation}
A different realization of this Heisenberg algebra promotes the coordinates 
(and thus all their functions) to noncommuting operators acting on an 
auxiliary Fock space~$\Hcal$ but keeps the ordinary operator product.
The Fock space is a Hilbert space with orthonormal basis states
\begin{equation} \label{oscbasis}
\begin{aligned}
{}& \qquad\qquad\qquad\quad |m\>\=\sfrac{1}{\sqrt{m!}}\,(\adag)^m\,|0\>
\qquad\text{for}\quad m\in\NN_0 \quad\text{and}\quad a|0\>=0 \ ,\\[4pt]
{}& a\,|m\> \= \sqrt{m}\,|m{-}1\> \ ,\qquad
\adag\,|m\> \= \sqrt{m{+}1}\,|m{+}1\> \ ,\qquad
N\,|m\> \ := \adag a\,|m\> \= m\,|m\> \ ,
\end{aligned}
\end{equation}
therewith characterizing $a$ and $a^\dagger$ as standard annihilation 
and creation operators. The star-product and operator formulations are 
tightly connected through the Moyal-Weyl map:
Coordinate derivatives correspond to commutators with coordinate operators, 
\begin{equation}
\sqrt{2\th}\,\pa_z \ \leftrightarrow\ -\text{ad}(\adag)\quad,\qquad
\sqrt{2\th}\,\pa_{\bar z} \ \leftrightarrow\ \text{ad}(a) \quad,
\end{equation}
and the integral over the noncommutative plane reads
\begin{equation}
\int\!\diff^2x\;f_{\star}(x) \= 2\pi\,\th\,\Tr_{\Hcal}\,f_{\mathrm{op}} \ ,
\end{equation}
where the function $f_{\star}$ corresponds to the operator~$f_{\mathrm{op}}$
via the Moyal-Weyl map and the trace is over the Fock space~$\Hcal$.
We shall work with the operator formalism but refrain from introducing
special notation indicating operators, so all objects are operator-valued
if not said otherwise.

\subsection{Two-dimensional sigma model}
\noindent
The fields $\Phi$ of the noncommutative two-dimensional U($n$) sigma model 
are unitary $n{\times}n$ matrices with operator-valued entries, 
i.e.~$\Phi\in\text{U}(\C^n\otimes\Hcal)=\text{U}(\Hcal^{\oplus n})$, 
and thus subject to the constraint
\begin{equation}
\Phi\,\Phi^\+ \= \unity_n\otimes\unity_{\Hcal} \= \Phi^\+\,\Phi \ .
\end{equation}
The Euclidean action of the model
coincides with the energy functional of its $2{+}1$ dimensional extension
evaluated on static configurations~\cite{Lechtenfeld:2001aw},\footnote{
$|A|^2 = \Tr(A^\+A)$ is the squared Hilbert-Schmidt norm of U($n$)-valued
operators on $\Hcal$. We restrict ourselves to finite-energy configurations,
i.e.~we demand that $[a,\Phi]$ exists and is Hilbert-Schmidt. Furthermore we
only consider solutions for which $\Delta\Phi$ is traceclass and $\Phi$ 
is a bounded operator in order for the below expressions to be well defined.}
\begin{equation} \label{E}
\begin{aligned}
E[\Phi] &\= 2\pi\,\th\,\Tr \bigl( 
\pa_z\Phi^\+\,\pa_{\bar z}\Phi + \pa_{\bar z}\Phi^\+\,\pa_z\Phi \bigr) \\
&\= \pi\,\Tr \bigl( [a,\Phi]^\+\,[a,\Phi] + [a,\Phi^\+] [a,\Phi^\+]^\+ \bigr)\\
&\= 2\,\pi\,\Tr \bigl( [a,\Phi]^\+\,[a,\Phi] \bigr)
 \= 2\,\pi\,\bigl| [a,\Phi] \bigr|^2 \\
&\= \pi\,\Tr \bigl( \Delta\Phi^\+\,\Phi + \Phi^\+\Delta\Phi \bigr) \ ,
\end{aligned}
\end{equation}
where the trace is taken over the Fock space $\Hcal$ as well as 
over the U($n$) group space.
Here, we have introduced the hermitian Laplace operator $-\Delta$
which is defined via
\begin{equation}
\Delta{\cal O} \ :=\ \bigl[a,[\adag,{\cal O}]\bigr]
\= \bigl[\adag,[a,{\cal O}]\bigr]\ .
\end{equation}
Its kernel is spanned by functions only of $a$ or only of $\adag$.
Varying $E[\Phi]$ under the above constraint one finds the equation of 
motion,\footnote{assuming that the appearing total derivative terms vanish}
\begin{equation} \label{fulleom}
0 \= \bigl[a\,,\Phi^\+[\adag,\Phi]\bigr] + \bigl[\adag,\Phi^\+[a\,,\Phi]\bigr]
  \= \Phi^\+ \, \Delta\Phi - \Delta\Phi^\+ \, \Phi \ .
\end{equation}

Since the annihilation and creation operators above should be read as
$\unity_n\otimes a$ and $\unity_n\otimes\adag$, respectively,
this model enjoys a global U(1)$\times$SU$(n)\times$SU($n$) invariance under
\begin{equation} \label{globalsym}
\Phi \quad\longrightarrow\quad 
(V\otimes\unity_{\Hcal})\,\Phi\,(W\otimes\unity_{\Hcal})
\end{equation}
with $V V^\+ = \unity_n = W W^\+$, which generates a moduli space
to any nontrivial solution of~(\ref{fulleom}). 
Another obvious symmetry is induced by the ISO(2) Euclidean group 
transformations of the noncommutative plane, 
as generated by the adjoint action of $a$, $\adag$ and $N$. 
Specifically, a global translation of the noncommutative plane,
\begin{equation}
(z\,,\,\bar z)\quad\longrightarrow\quad
(z+\zeta\,,\,\bar z+\bar\zeta)\=\sqrt{2\th}\,(a+\a\,,\,\adag+\ab)\ ,
\end{equation}
induces on (operator-valued) scalar functions~$f$ the unitary operation
\begin{equation} \label{globaltrans}
f\quad\longrightarrow\quad \e^{-\zeta\pa_z-\bar\zeta\pa_{\bar z}}\,f 
\= \e^{\a\,\text{ad}(\adag)-\ab\,\text{ad}(a)}\,f
\= \e^{\a\adag-\ab a}\,f\,\e^{-\a\adag+\ab a}\ =:\ D(\a)\,f\,D(\a)^\+\ .
\end{equation}
Its action on states leads to coherent states, e.g.
\begin{equation} \label{coherent}
D(\a)\,|0\> \= \e^{\a\adag-\ab a}\,|0\> 
\= \e^{-\frac12\ab\a}\,\e^{\a\adag}\,|0\> \ =:\ |\a\> \ .
\end{equation}
A global rotation of the noncommutative plane,
\begin{equation}
(z\,,\,\bar z)\quad\longrightarrow\quad 
(\e^{\ic\,\vartheta} z\,,\,\e^{-\ic\,\vartheta} \bar z)
\= \sqrt{2\th}\,(\e^{\ic\,\vartheta} a\,,\,\e^{-\ic\,\vartheta} \adag)\ ,
\end{equation}
induces the unitary transformation
\begin{equation} \label{globalrot}
f\quad\longrightarrow\quad \e^{\ic\vartheta\,\text{ad}(\adag a)}\,f
\= \e^{\ic\vartheta\,\adag a}\,f\,\e^{-\ic\vartheta\,\adag a}
\ =:\ R(\vartheta)\,f\,R(\vartheta)^\+
\end{equation}
for $\vartheta\in\R/2\pi\Z$, 
because $[N,f]=\bar{z}\pa_{\bar z}f-\pa_z f z$.    
Applying $R(\vartheta)$ to a coherent state we obtain
\begin{equation} 
R(\vartheta) |\a\> \= |\e^{\ic\,\vartheta} \a\>\ .
\end{equation}
Since the adjoint actions of both~$D(\a)$ and~$R(\vartheta)$ commute with 
that of~$\Delta$, the energy functional is invariant under them,
and all translates and rotations of a solution to~(\ref{fulleom}) 
also qualify as solutions, with equal energy. 
However, other unitary transformations will in general change the value of~$E$.

There is a wealth of finite-energy configurations~$\Phi$ which fulfil 
the equation of motion~(\ref{fulleom}).
A subclass of those is distinguished by possessing a smooth commutative limit,
in which $\Phi$ merges with a commutative solution. The latter have been
classified by \cite{Uhlenbeck, Wood}. We call these ``nonabelian'' because they
cannot appear in the U(1) case, where the commutative model is a free field 
theory and does not admit finite-energy solutions.
Yet, there can (and do) exist non-trivial abelian finite-energy solutions
at finite~$\th$, whose $\th\to0$ limit is necessarily singular and
produces a discontinuous configuration. We term such solutions ``abelian''
even in case they are imbedded in a nonabelian group.

Of particular interest are
configurations diagonal in the oscillator basis~(\ref{oscbasis}) as well as
in~$\C^n$, namely $\Phi=
\bigoplus_{i=1}^n\text{diag}\bigl(\{\e^{\ic\a_i^\ell}\}_{\ell=0}^\infty\bigr)$
with $\a_i^\ell\in\R$. 
A short computation reveals that such a configuration
obeys the equation of motion~(\ref{fulleom}) if and only if the phases
satisfy $\e^{\ic\a_i^\ell}=\pm\e^{\ic\a_i}$ where the sign depends on~$\ell$.
Its energy is finite if the number of positive signs or the number of
negative signs in each block labelled by~$i$ is finite.

\subsection{Grassmannian configurations}
\noindent
It is a daunting task to classify all solutions to the full equation of
motion~(\ref{fulleom}), and we shall focus on the subset of hermitian ones.
Any (not necessarily classical) hermitian configuration obeys
\begin{equation}
\Phi^\+ \= \Phi \qquad\Longrightarrow\qquad 
\Phi^2 \= \unity_n\otimes\unity_{\Hcal} \ =:\ \unity \ ,
\end{equation}
and is conveniently parametrized by a hermitian projector~$P{=}P^\+$ via
\begin{equation}
\Phi \ =:\ \unity - 2P \= P^\perp - P\= \e^{\ic\pi P}
\qquad\text{with}\quad P^2 = P\ ,
\end{equation}
where $P^\perp=\unity{-}P$ denotes the complementary projector.
This allows one to define a topological charge like in the commutative case as
\cite{Matsuo:2000pj}
\begin{equation} \label{Qdef}
\begin{aligned}
Q[\Phi] &\= \sfrac{1}{4}\,\th\,\Tr \bigl(
\Phi\,\pa_z\Phi\,\pa_{\bar z}\Phi - \Phi\,\pa_{\bar z}\Phi\,\pa_z\Phi \bigr)\\
&\= \Tr \bigl( P\,[\adag,P]\,[a\,,P]-P\,[a\,,P]\,[\adag,P] \bigr) \\
&\= \bigl| P\,[a\,,P] \bigr|^2 - \bigl| [a\,,P]\,P \bigr|^2 \\
&\= \Tr \bigl( P - P\,a\,P\,\adag P + P\,\adag P\,a\,P \bigr) \\
&\= \Tr \bigl(P\,a\,(\unity{-}P)\,\adag P-P\,\adag(\unity{-}P)\,a\,P\bigr)\ ,
\end{aligned}
\end{equation}
which may be compared with the energy
\begin{equation} \label{Edef}
\begin{aligned}
\sfrac{1}{8\pi}\,E[\Phi] &\= \sfrac12\,\th\,\Tr \bigl(
\pa_z \Phi\,\pa_{\bar z}\Phi \bigr) \\
&\= \Tr \bigl( [\adag,P]\;[P\,,a] \bigr) \\
&\= \bigl| P\,[a\,,P] \bigr|^2 + \bigl| [a\,,P]\,P \bigr|^2 \\
&\=\Tr\bigl( P\,(a\,\adag{+}\adag a)\,P
           - P\,a\,P\,\adag P - P\,\adag P\,a\,P \bigr) \\
&\= \Tr \bigl(P\,a\,(\unity{-}P)\,\adag P+P\,\adag(\unity{-}P)\,a\,P\bigr)\ .
\end{aligned}
\end{equation}

Each hermitian projector~$P$ gives rise to the complementary projector
$\unity{-}P$, with the properties\footnote{
By a slight abuse of notation, we denote the energy and topological charge
as functionals of~$P$ again with the symbols $E$ and~$Q$, respectively.}
\begin{equation} \label{coP}
E[\unity{-}P] \= E[P] \qquad\text{and}\qquad
Q[\unity{-}P] \= -Q[P] \ .
\end{equation}
Comparing (\ref{Qdef}) and (\ref{Edef}) we get the relations
\begin{equation} \label{EandQ}
\sfrac{1}{8\pi}\,E[P] \= Q[P] + 2 \bigl| [a\,,P]\,P \bigr|^2
\= -Q[P] + 2 \bigl| P\,[a\,,P] \bigr|^2 \ ,
\end{equation}
which yield the BPS bound
\begin{equation} \label{bound}
E[P]\ \ge\ 8\pi\,\bigl| Q[P] \bigr| 
\end{equation}
for any hermitian configuration.

The set of all projectors unitarily equivalent to $P$ is called
the Grassmannian, and it is given by the coset
\begin{equation} \label{defGr}
\text{Gr}(P) \= 
\frac{\text{U}(\Hcal^{\oplus n})}{\text{U(im$P$)}\times\text{U(ker$P$)}}\ .
\end{equation}
Thus, each given $P$ (and thus hermitian $\Phi$) belongs to a certain
Grassmannian, and the space of all hermitian~$\Phi$ decomposes into a
disjoint union of Grassmannians. The restriction to hermitian $\Phi$ 
reduces the unitary to a Grassmannian sigma model, whose configuration space 
is parametrized by projectors $P$.
In the commutative case the topological charge $Q$ is an element of
$\pi_2(\text{Gr}(P))=\Z$, where $S^2$ is the compactified plane. Hence, 
it is an invariant of the Grassmannian if one excludes from (\ref{defGr})
any singular unitary transformations with nontrivial winding at infinity.
It is less obvious how to properly extend this consideration to the
infinite-dimensional cases encountered here~\cite{Harvey:2001pd}. 
We therefore take a pragmatic viewpoint and demand $Q[P]$ to be constant 
throughout the Grassmannian. This may downsize the above coset $\text{Gr}(P)$ 
by restricting the set of admissible unitaries~$U$. 

Let us make this more explicit by computing $Q[U\!PU^\+]$.
To this end, we define
\begin{equation} \label{defomega}
\omega\ :=\ U^\+\,[a\,,\,U] \qquad\text{and}\qquad
\omega^\+\= U^\+\,[\adag,\,U] \ ,\qquad\text{with}\qquad
[a+\omega\,,\adag+\omega^\+]\=\unity \ ,
\end{equation}
as elements of the Lie algebra of U$(\Hcal^{\oplus n})$.\footnote{
In fact, the finite-energy condition (see footnote before (\ref{E})) 
enforces $[a\,,U\!PU^\+]$ to be Hilbert-Schmidt which implies that
$[\omega\,,P]$ is Hilbert-Schmidt as well. Unfortunately, this does not
suffice to guarantee the constancy of $Q$ in $\text{Gr}(P)$. 
On the other hand, $\omega$ itself need not even be bounded, as in the 
example of (\ref{globalrot}) where $\omega=(\e^{\ic\vartheta}{-}1)a$.}
A short calculation yields
\begin{align} \label{QU}
Q[U\!PU^\+] &\= \Tr \bigl( 
P\,[\adag{+}\omega^\+,P]\,[a{+}\omega\,,P] -
P\,[a{+}\omega\,,P]\,[\adag{+}\omega^\+,P] \,\bigr) \\[4pt] 
&\= Q[P]\;+\;\Tr\,P\,\bigl( \,
[\adag\!,P]\omega + \omega[\adag\!,P] - [a,P]\omega^\+\! - \omega^\+[a,P]
- \omega^\+(\unity{-}P)\omega + \omega(\unity{-}P)\omega^\+ \bigr) P \ , 
\nonumber
\end{align}
which constrains $\omega$ in terms of~$P$. In case of $\Tr P<\infty$,
this indeed reduces to
\begin{equation}
Q[U\!PU^\+] \= Q[P]\ +\ \Tr\,P\,\bigl( \,
[\omega\,,\adag] + [a\,,\omega^\+] + [\omega\,,\omega^\+] \bigr) \= Q[P] \ .
\end{equation}

Alternatively, we may calculate the infinitesimal variation of $Q[P]$
under $P\to P{+}\de P$. Remembering that $P$ and $\de P$ are bounded and that
their commutators with $a$ or $\adag$ are Hilbert-Schmidt, we get
\begin{equation} \label{Qvar}
\de Q[P] \= \Tr \bigl( 
\bigl[\adag,[P,\de P][a,P]\bigr]-\bigl[a,[P,\de P][\adag,P]\bigr] \bigr)
\ +\ 3\,\Tr \,\de P \bigl( [\adag,P][a,P] - [a,P][\adag,P] \bigr)
\end{equation}
with the first trace being a ``boundary term''.
Variations inside the Grassmannian are given by
\begin{equation}
\de P \= [ \Lambda_{\text{o}}\,,P ] 
\qquad\text{with}\qquad \Lambda_{\text{o}}^\+ = -\Lambda_{\text{o}} 
\qquad\text{and}\qquad 
P\Lambda_{\text{o}}P \=0\= (\unity{-}P)\Lambda_{\text{o}}(\unity{-}P)
\end{equation}
(see Section~3.2 below), and thus
\begin{equation}
\de Q[P] \= \Tr \bigl(
\bigl[a , \Lambda_{\text{o}} [\adag,P]\bigr] -
\bigl[\adag , \Lambda_{\text{o}} [a,P]\bigr] \bigr) 
\ +\ 3\,\Tr \, \bigl[ P\,,\,
\Lambda_{\text{o}} [a,P][\adag,P] - \Lambda_{\text{o}} [\adag,P][a,P] \bigr]\ .
\end{equation}
Hence, if $\Lambda_{\text{o}}$ is bounded, the second term vanishes
and $\de Q[P]$ reduces to the boundary term.\footnote{
Yet again, this condition is too strong a demand as the examples of 
(\ref{globaltrans}) and (\ref{globalrot}) indicate.}

Formally, another invariant is the rank of the projector,
\begin{equation}
r \ :=\ \text{rank}(P) \= \text{dim(im}P) \ ,
\end{equation}
which may differ from $Q$ when being infinite.
Let us consider the class of projectors which can be decomposed as
\begin{equation} \label{specialP}
P \= U\,\bigl( \widehat{P}\,+\,P' \bigr) \, U^\+ \qquad\text{with}\qquad
\widehat{P} = \bar{P}\otimes\unity_\Hcal \qquad\text{and}\qquad
\widehat{P}\,P' =0= P'\,\widehat{P} \ ,
\end{equation}
where $U$ is an admissible unitary transformation, $\bar{P}$ denotes a
constant projector, and the rank~$r'$ of $P'$ is finite.
In this case the difference of $r$ and $Q$ is determined by the invariant
U($n$) trace~\cite{Otsu}
\begin{equation}
R[P] \= \tr\,\bar P\ \in \{0,1,\dots,n\}\ ,
\end{equation}
and such projectors are characterized by the pair $(R,Q)$.
Formally, the total rank then becomes $r=R{\cdot}\infty+Q$.
Since $a$ and $\adag$ commute with $\widehat{P}$,
the topological charge depends only on the finite-rank piece,
\begin{equation} \label{specialQE}
Q[P] \= Q[\widehat{P}{+}P'] \= Q[P'] \= \Tr\,P' \= r' \ .
\end{equation}
For $\bar P{=}0$ one obtains the abelian solutions because they fit 
into~U($\Hcal$). 
For $R{>}0$ we have nonabelian solutions because more than one copy of
$\Hcal$ is needed to accomodate them.

Any projector in $\C^n\otimes\Hcal=\Hcal^{\oplus n}$ can be parametrized as
\begin{equation} \label{Tdef}
P \= |T\>\,\<T|T\>^{-1}\<T|
\end{equation}
where 
\begin{equation} \label{Tarray}
|T\> \= 
\Bigl(\, |T^1\> \quad |T^2\> \quad \dots \quad |T^r\> \Bigr) \=
\begin{pmatrix}
|T_1^1\> & |T_1^2\> & \dots  & |T_1^r\> \\[4pt]
|T_2^1\> & |T_2^2\> & \dots  & |T_2^r\> \\[4pt]
\vdots   & \vdots   & \ddots & \vdots   \\[4pt]
|T_n^1\> & |T_n^2\> & \dots  & |T_n^r\> \end{pmatrix} \=
\begin{pmatrix}
|T_1\> \\[4pt] |T_2\> \\[4pt] \vdots \\[4pt] |T_n\> \end{pmatrix} 
\end{equation}
denotes an $n{\times}r$ array of kets in $\Hcal$,
with $r$ possibly being infinite. Thus,
\begin{equation}
\<T|T\> \= \Bigl( \<T^\ell|T^m\> \Bigr) \=
\Bigl( \textstyle{\sum_{i=1}^n} \<T_i^\ell|T_i^m\> \Bigr)
\end{equation}
stands for an invertible $r{\times}r$ matrix, and $r$ is the rank of~$P$.
The column vectors $|T^\ell\>$ span the image im$P$ of the projector in
$\Hcal^{\oplus n}$.
There is some ambiguity in the definition~(\ref{Tdef}) of~$|T\>$ since
\begin{equation}
|T\>\ \to\ |T\>\,\Gamma 
\qquad\text{for}\quad \Gamma\in\text{GL}(r)
\end{equation}
amounts to a change of basis in im$P$ and does not change the projector~$P$.
This freedom may be used to normalize $\<T|T\>=\unity_r$, which is still
compatible with $\Gamma\in\text{U}(r)$.
On the other hand, a unitary transformation
\begin{equation}
|T\>\ \to\ U\,|T\> \qquad\text{for}\quad U\in\text{U}(\Hcal^{\oplus n})
\end{equation}
yields a unitarily equivalent projector $UP\,U^\+$.
Altogether, we have the bijection 
\begin{equation} \label{Ttransform}
\tilde{P} \= U\,P\,U^\+ \qquad\Longleftrightarrow\qquad
|\tilde{T}\> \= U\,S\,|T\> \= U\,|T\>\,\Gamma 
\qquad\text{with}\quad S \in \text{GL(im$P$)} \ .
\end{equation}
Here, the trivial action of $U\in\text{U(im$P$)}{\times}\text{U(ker$P$)}$
on $P$ can be subsumed in the $S$ action on~$|T\>$.

Any diagonal projector can be cast into the form
\begin{equation} \label{Pdiag}
P_{\text{d}} \= \text{diag}\,\Bigl( 
\underbrace{\unity_\Hcal,\dots,\unity_\Hcal}_{R\text{\ times}}\,,\,P_Q\,, 
\underbrace{{\bf0}_\Hcal,\dots,{\bf0}_\Hcal}_{n{-}R{-}1\text{\ times}} \Bigr)
\qquad\text{with}\qquad P_Q \= \sum_{m=0}^{Q-1} |m\>\<m| \ .
\end{equation}
Formally, $P_{\text{d}}$ has rank~$r$ in $\Hcal^{\oplus n}$. 
A corresponding $r{\times}n$ array of kets via (\ref{Tdef}) would be
\begin{equation} \label{specialket}
|T_{\text{d}}\> \=
\begin{pmatrix}
|\Hcal\>  & \emptyset & \dots  & \emptyset & 0_Q    \\
\emptyset & |\Hcal\>  &        & \emptyset & 0_Q    \\
\vdots    &           & \ddots &           & \vdots \\
\emptyset & \emptyset &        & |\Hcal\>  & 0_Q    \\
\emptyset & \emptyset & \dots  & \emptyset & |T_Q\> \\
\vdots    & \vdots    &        & \vdots    &        \\
\emptyset & \emptyset & \dots  & \emptyset & 0_Q
\end{pmatrix}
\qquad\text{with}\quad\begin{cases} {}\\[-12pt]
\quad|\Hcal\> &\!\!=\ \bigl( |0\>\ |1\>\ |2\>\ |3\>\ \dots \bigr) \\[4pt] {}\
\quad\emptyset &\!\!=\ \bigl(\ 0\quad0\quad0\quad0\ \ \dots \bigr) \\[4pt]
\quad|T_Q\> &\!\!=\ \bigl( |0\>\ |1\>\ \dots\ |Q{-}1\> \bigr) \\[4pt] {}\
\quad0_Q &\!\!=\ 
    \bigl(\ \underbrace{0\quad0\quad\dots\quad0}_{Q\text{\ times}}\ \bigr)
\end{cases} \quad.
\end{equation}
In the abelian case, $n{=}1$ and $R{=}0$, 
this reduces to $|T_{\text{d}}\>=|T_Q\>$, 
and any projector is unitarily equivalent to~$P_Q$.
Due to (\ref{coP}),
analogous results are valid in the complementary case $P\to\unity{-}P$.

\subsection{BPS solutions}
\noindent
Let us focus on classical solutions within a Grassmannian.
For $\Phi=\unity{-}2P$, the equation of motion~(\ref{fulleom}) reduces to
\begin{equation} \label{Peom}
0 \= [ \Delta P\,,\,P ] \=
[ \adag,\,(\unity{-}P)a P ] + [ a\,,\,P \adag(\unity{-}P) ] \=
[ a\,,\,(\unity{-}P)\adag P ] + [ \adag,\,P a(\unity{-}P) ] \ .
\end{equation}
Still, we do not know how to characterize its full solution space.
However, (\ref{Peom}) is identically satisfied by projectors subject to
\cite{Gopakumar:2001yw,Hadasz:2001cn}
\begin{align}
\text{either the BPS equation}&& 
0 &\= [a\,,P]\,P \= (\unity{-}P)\,a\,P &\label{BPS}\\
\text{or the anti-BPS equation}&& 
0 &\= [\adag,P]\,P \= (\unity{-}P)\,\adag P &\label{aBPS}
\end{align}
which are only ``first-order''.
Solutions to (\ref{BPS}) are called solitons while those to (\ref{aBPS}) are
named anti-solitons. Hermitian conjugation shows that the latter are obtained 
{}from the former by exchanging $P\leftrightarrow\unity{-}P$, and so we can 
ignore the anti-BPS solutions for most of the paper. 
Equation (\ref{BPS}) means that
\begin{equation}
a \quad\text{maps}\quad \text{im}P\ \hookrightarrow\ \text{im}P
\end{equation}
and, hence, characterizes subspaces of $\Hcal^{\oplus n}$ which are
stable under the action of~$a$.
The term ``BPS equation'' derives from the observation that (\ref{BPS})
inserted in (\ref{EandQ}) implies the saturation of the BPS 
bound~(\ref{bound}), which simplifies to
\begin{equation}
E[P] \= 8\pi\,Q[P] \= 
8\pi\,\Tr \bigl( P\,a\,(\unity{-}P)\,\adag P \bigr) \=
8\pi\,\Tr \bigl( P\,a\,\adag P - a\,P\,\adag \bigr) \ .
\end{equation}
For BPS solutions of the particular form (\ref{specialP}) we indeed
recover that $E[P]=8\pi\,\Tr P'$.
Clearly, these BPS solutions constitute the absolute minima of the energy 
functional within each Grassmannian.

When the parametrization (\ref{Tdef}) is used, the BPS condition (\ref{BPS}) 
simplifies to
\begin{equation} \label{BPS2}
a\,|T_i^\ell\> \= |T_i^{\ell'}\>\,\gamma_{\ell'}^{\ \ell} 
\qquad\text{for some $r{\times}r$ matrix}\quad 
\gamma\=\bigl(\gamma_{\ell'}^{\ \ell}\bigr)
\end{equation}
which represents the action of~$a$ in the basis chosen for im$P$.
For instance, the ket $|T_Q\>$ in (\ref{specialket}) indeed obeys
\begin{equation}
a\,|T_Q\> \= |T_Q\>\,\gamma_Q \qquad\text{with}\qquad \gamma_Q \=
\left( \begin{smallmatrix}
0      & \sqrt{1} & 0        & \dots  & 0            \\
0      & 0        & \sqrt{2} &        & 0            \\
\vdots & \vdots   & \ddots   & \ddots &              \\
0      & 0        & \dots    & 0      & \sqrt{Q{-}1} \\
0      & 0        & \dots    & 0      & 0
\end{smallmatrix} \right)
\quad,
\end{equation}
and the diagonal projector $P_{\text{d}}$ in (\ref{Pdiag}) is BPS,
but it is by far not the only one.

Any basis change in im$P$ induces a similarity transformation
$\gamma\mapsto\Gamma\gamma\Gamma^{-1}$, which leaves $P$ unaltered and
thus has no effect on the value of the energy. Therefore, it suffices
to consider $\gamma$ to be of Jordan normal form. More generally,
a unitary transformation~(\ref{Ttransform}) is compatible with the 
BPS condition (\ref{BPS2}) only if 
\begin{equation}
U^\+ a\,U\,|T\> \= |T\>\,\gamma_U 
\qquad\text{for some $r{\times}r$ matrix}\ \gamma_U \ ,
\end{equation}
which implies that
\begin{equation} \label{BPScomp}
\omega\,|T\> \= |T\>\,\gamma_\omega 
\qquad\text{with}\qquad \gamma_\omega = \gamma_U-\gamma \ .
\end{equation}
The trivially compatible transformations are those in
$\text{U(im$P$)}{\times}\text{U(ker$P$)}$, which can be subsumed in
$S\in\text{GL(im$P$)}$ and lead to $\gamma_U{=}\Gamma^{-1}\gamma\Gamma$.
Another obvious choice are rigid symmetries of the energy functional, 
as given in (\ref{globaltrans}), (\ref{globalrot}), and (\ref{globalsym}) 
for $W{=}V^\+$.
The challenging task then is to identify the nontrivial BPS-compatible unitary 
transformations, since these relate different $a$-stable subspaces of fixed 
dimension in $\Hcal^{\oplus n}$ and thus generate the multi-soliton moduli 
space. For $|T_Q\>$ in (\ref{specialket}) we shall accomplish this 
infinitesimally in Section~3.3.

\subsubsection{Abelian solitons}
\noindent
Let us take a closer look at the BPS solutions of the noncommutative
U(1) sigma model. All finite-energy configurations are based on $R{=}0$
and have rank $r=Q<\infty$, thus $E=8\pi r$. 
The task is to solve the ``eigenvalue equation''
\begin{equation} \label{BPSabelian}
a\,|T\> \= |T\>\,\gamma \qquad\text{for}\qquad \gamma \= \bigoplus_{s=1}^q
\left( \begin{smallmatrix}
\a_s   & 1      & 0      & \dots  & 0    \\
0      & \a_s   & 1      &        & 0    \\
\vdots &        & \ddots & \ddots &      \\
0      & 0      &        & \a_s   & 1    \\
0      & 0      & \dots  & 0      & \a_s 
\end{smallmatrix} \right)  \qquad\text{with}\quad \a_s\in\C \ ,
\end{equation}
where the Jordan cells have sizes~$r_s$ for $s=1,\dots,q$, 
with $\sum_{s=1}^q r_s=r$.
For a given rank~$r$, the above matrices $\gamma$ parametrize the 
$r$-soliton moduli space.
The general solution (unique up to cell-wise normalization and basis
changes $|T^{(s)}\>\to|T^{(s)}\>\Gamma^{(s)}$) reads
\begin{equation}
|T\> \= \Bigl( |T^{(1)}\>\ \dots\ |T^{(q)}\> \Bigr)
\qquad\text{with}\qquad |T^{(s)}\> \= \Bigl( 
|\a_s\>\ \ a^\+|\a_s\>\ \dots\ \ \sfrac{1}{(r_s-1)!}(a^\+)^{r_s-1}|\a_s\>
\Bigr)
\end{equation}
and is based on the coherent states~(\ref{coherent}). 
In the star-product picture, the corresponding $\Phi$ represents $r$ lumps
centered at positions $\a_s$ with degeneracies $r_s$ in the $xy$~plane.
Lifting a degeneracy by ``point-splitting'', the related Jordan cell dissolves
into different eigenvalues. Hence, the generic situation has
$r_s{=}1\ \forall s$, and so
\begin{equation} \label{Tcoherentr}
\gamma \= \text{diag} (\a_1,\a_2,\dots,\a_r) \qquad\Longleftrightarrow\qquad
|T\> \= \Bigl( |\a_1\>\ |\a_2\>\ \dots\ |\a_r\> \Bigr) \ ,
\end{equation}
which yields the projector
\begin{equation} \label{Pcoherentr}
P \= \sum_{k,\ell=1}^r 
|\a_k\> \,\Bigl( \<\a_.|\a_.\> \Bigr)^{-1}_{k\ell} \<\a_\ell| \ .
\end{equation}

Each solution can be translated via a unitary transformation mediated
by $D(\beta)$, which shifts $\a_\ell\mapsto\a_\ell{+}\beta\ \forall\ell$,
and rotated by $R(\vartheta)$, which moves 
$\a_\ell\mapsto\e^{\ic\vartheta}\a_\ell\ \forall\ell$.
The individual values of~$\a_\ell$ (the soliton locations) may also be moved 
around by appropriately chosen unitary transformations, 
so that any $r$-soliton configuration can be reached from the diagonal one,
which describes $r$ solitons on top of each other at the coordinate origin
\cite{Hadasz:2001cn}:
\begin{equation} \label{diagonalBPS}
|T\> \= U\,|T_r\>\,\Gamma \qquad\Longleftrightarrow\qquad
P \= U\,P_r\,U^\+ \qquad\text{with}\quad 
P_r\= \sum_{m=0}^{r-1} |m\>\<m| \ .
\end{equation}
We illustrate the latter point with the example of $r=2$.  Generically, 
\begin{equation}
\bigl( |\a_1\>\ |\a_2\> \bigr) \= U\,S\,\bigl( |0\>\ |1\> \bigr) 
\= U\,\bigl( |0\>\ |1\> \bigr)\,\Gamma \qquad\text{with}\quad
U\,S\, \= |\a_1\>\<0| + |\a_2\>\<1| + \dots \ ,
\end{equation}
where the omitted terms annihilate $|0\>$ and $|1\>$. Factorizing $US$ yields
\begin{equation} \label{U2}
\begin{aligned}
U &\= \frac{1}{\sqrt{1{-}|\sigma|^2}}\,\Bigl(|\a_1\>\ |\a_2\>\Bigr)
\begin{pmatrix} \e^{-\ic\gamma} & 0 \\ 0 & \e^{-\ic\gamma'} \end{pmatrix}
\begin{pmatrix} -\sin\beta' & \phantom{-}\cos\beta' \\ 
                \phantom{-}\sin\beta & -\cos\beta \end{pmatrix}
\begin{pmatrix} \<0| \\ \<1| \end{pmatrix} \ +\ \dots  \\[4pt]
\text{and}\qquad \Gamma &\= 
\begin{pmatrix} \cos\beta & \cos\beta' \\ \sin\beta & \sin\beta' \end{pmatrix}
\begin{pmatrix} \e^{\ic\gamma} & 0 \\ 0 & \e^{\ic\gamma'} \end{pmatrix} 
\ ,\\[4pt]  \text{with}\qquad \sigma &\= 
\<\a_1|\a_2\> \= \e^{\bar{\a}_1\a_2-\frac12|\a_1|^2-\frac12|\a_2|^2}
\= \e^{-\ic(\gamma-\gamma')} \cos(\beta{-}\beta') \ ,
\end{aligned}
\end{equation}
so that $|\sigma|^2=\e^{-|\a_1-\a_2|^2}$, and $\beta{+}\beta'$ and
$\gamma{+}\gamma'$ remain undetermined. The rank-2 projector becomes
\begin{equation}
P \= \frac{1}{1{-}|\sigma|^2}\,\Bigl( |\a_1\>\<\a_1| + |\a_2\>\<\a_2|
- \sigma |\a_1\>\<\a_2| - \bar\sigma |\a_2\>\<\a_1| \Bigr) \=
U\, \Bigl( |0\>\<0| + |1\>\<1| \Bigr)\, U^\+ \ .
\end{equation}

\subsubsection{Nonabelian solitons}
\noindent
We proceed to infinite-rank projectors. For simplicity, let us discuss
the case of U(2) solitons -- the results will easily generalize to U($n$).
Clearly, we can imbed two finite-rank BPS solutions (with $R{=}0$) into 
U($\Hcal{\oplus}\Hcal$) by letting each act on a different copy of~$\Hcal$.
Such configurations are noncommutative deformations of the trivial projector 
$\bar{P}=(\begin{smallmatrix} 0&0\\0&0 \end{smallmatrix})$ 
and thus represent a combination of abelian solitons. Therefore, we turn to 
projectors with $R=1$ and so, formally, $r=\infty+Q$. Given (\ref{specialP})
such solutions may be considered as noncommutative deformations of the
U(2) projector $\bar{P}=(\begin{smallmatrix} 1&0\\0&0 \end{smallmatrix})$.
For this reason, one expects the generic solution $|T\>$ to (\ref{BPS2})
to be combined from a full set of states for the first copy of $\Hcal$
and a finite set of coherent states $|\a_\ell\>$ in the second copy of $\Hcal$, 
\begin{equation}
|T\> \= \begin{pmatrix} 
|\Hcal\>  & 0       & 0       & \dots & 0 \\[4pt]
\emptyset & |\a_1\> & |\a_2\> & \dots & |\a_Q\> \end{pmatrix}
\qquad\Longrightarrow\qquad
P \= \unity_\Hcal\ \oplus\ \sum_{k,\ell=1}^Q 
|\a_k\> \,\Bigl( \<\a_.|\a_.\> \Bigr)^{-1}_{k\ell} \<\a_\ell| \ ,
\end{equation}
with $E=8\pi Q$. 
This projector is of the special form (\ref{specialP}), with
$\bar{P}=(\begin{smallmatrix}1&0\\0&0\end{smallmatrix})$ and $U=\unity$.
By a unitary transformation on the second copy of~$\Hcal$ 
such a configuration can be mapped to the diagonal form
\begin{equation} \label{simplespecialP} 
|T_{\text{d}}\> \= \begin{pmatrix} 
|\Hcal\>  & 0    & 0    & \dots & 0 \\[4pt]
\emptyset & |0\> & |1\> & \dots & |Q{-}1\> \end{pmatrix}
\qquad\Longrightarrow\qquad
P_{\text{d}} \= \unity_\Hcal\ \oplus\ P_Q\ .
\end{equation}
It is convenient to reorder the basis of im$P_{\text{d}}$ such that
\begin{equation} \label{Thatdef}
|T_{\text{d}}\> \= \begin{pmatrix} 
0    & 0    & \dots & 0        & |\Hcal\> \\[4pt]
|0\> & |1\> & \dots & |Q{-}1\> & \emptyset \end{pmatrix}
\= \begin{pmatrix} S_Q \\[4pt] P_Q \end{pmatrix} |\Hcal\> \ 
=:\ \hat{T}_{\text{d}}\,|\Hcal\>\ ,
\end{equation}
where
\begin{equation}
P_Q \= \sum_{m=0}^{Q-1} |m\>\<m| \qquad\text{and}\qquad
S_Q \= (a\,\sfrac{1}{\sqrt{N}})^Q \= \sum_{m=Q}^\infty |m{-}Q\>\<m|
\end{equation}
denotes the $Q$th power of the shift operator.
The form of (\ref{Thatdef}) suggests to pass from states 
$|T\>=\begin{pmatrix} |T_1\> \\ |T_2\> \end{pmatrix}$ with $R{=}1$ 
to operators $\hat{T}=\begin{pmatrix} \hat{T}_1 \\ \hat{T}_2 \end{pmatrix}$
on $\Hcal$:
\begin{equation}
|T\> \= \hat{T}\,|\Hcal\> \qquad\Longrightarrow\qquad
P \= \hat{T}\,(\hat{T}^\+ \hat{T})^{-1} \hat{T}^\+ \ .
\end{equation}
In fact, it is always possible to introduce $\hat{T}$ as
\begin{equation}
\hat{T}_i \= \sum_{\ell=1}^r |T_i^\ell\>\<\ell{-}1| 
\qquad\Longleftrightarrow\qquad 
|T_i^\ell\> \= \hat{T}_i\,|\ell{-}1\> 
\qquad \textrm{for} \quad i=1,2 \quad\textrm{and}\quad \ell=1,\dots,r\ .
\end{equation}
We may even put $\hat{T}^\+\hat{T}=\unity_\Hcal$ by using the
freedom $\hat{T}\to\hat{T}\,\hat\Gamma$ with an operator 
$\hat\Gamma=(\hat{T}^\+\hat{T})^{-1/2}$. Our example 
of $|T_{\text{d}}\>$ in~(\ref{Thatdef}) is already normalized since
\begin{equation}
S_1\,S_1^\+\=\unity_\Hcal \qquad\text{but}\qquad 
S_1^\+ S_1 \=\unity_\Hcal-|0\>\<0|
\qquad\Longrightarrow\qquad S^\+_Q S_Q + P_Q \= \unity_\Hcal \ .
\end{equation}

It is instructive to turn on a BPS-compatible unitary transformation
in~(\ref{specialP}). In our example~(\ref{Thatdef}), we apply~\cite{Lee:2004dt}
\begin{equation} \label{Umu}
U(\mu) \= \begin{pmatrix}
S_Q \sqrt{\frac{N_Q}{N_Q{+}\mu\bar\mu}} \, S^\+_Q &
S_Q \frac{\bar\mu}{\sqrt{N_Q{+}\mu\bar\mu}} \\[12pt]
\frac{\mu}{\sqrt{N_Q{+}\mu\bar\mu}} \, S^\+_Q &
\frac{\mu\,P_Q\,-\,\sqrt{N_Q}}{\sqrt{N_Q{+}\mu\bar\mu}}
\end{pmatrix} 
\qquad\text{with}\quad
N_Q \= {\adag}^Q a^Q \= N(N{-}1)\cdots(N{-}Q{+}1) 
\end{equation}
to $\ \hat{T}_{\text{d}}=(S_Q,P_Q)^t\ $ of (\ref{Thatdef}). With the help of
\begin{equation}
S_Q P_Q \= 0 \= N_Q P_Q \qquad\text{and}\qquad
S_Q \sqrt{N_Q} \= a^Q
\end{equation}
we arrive at
\begin{equation} \label{Ttilde}
\hat{T}(\mu) \= U(\mu)\,\hat{T}_{\text{d}} \=
\begin{pmatrix} a^Q \\ \mu \end{pmatrix} \,
\frac{1}{\sqrt{N_Q{+}\mu\bar\mu}} \= 
\begin{pmatrix} a^Q \\ \mu \end{pmatrix} \,\hat\Gamma 
\ =:\ \check{T}(\mu)\,\hat\Gamma \ .
\end{equation}
This transformation can be regarded as a regularization 
of $\hat{T}_{\text{d}}$ since
\begin{equation}
\lim_{\mu\to0} \hat{T}(\mu) \= 
\bigl(\begin{smallmatrix} 1 & 0 \\ 0 & \e^{\ic\delta}\!\end{smallmatrix}\bigr) 
\,\hat{T}_{\text{d}} \quad\text{with}\quad 
\delta=\lim_{\mu\to0}\arg\mu \qquad\text{and thus}\qquad
\lim_{\mu\to0} P(\mu) \= P_{\text{d}} \ ,
\end{equation}
but note the singular normalization in the limit! For completeness we also
display the transformed projector,
\begin{equation}
P(\mu) \= U(\mu) \begin{pmatrix} 
\unity_\Hcal & {\bf0}_\Hcal \\[8pt] {\bf0}_\Hcal & P_Q 
\end{pmatrix} U^\+(\mu) \= \begin{pmatrix}
a^Q \frac{1}{N_Q{+}\mu\bar\mu} {\adag}^Q & 
a^Q \frac{\bar\mu}{N_Q{+}\mu\bar\mu} \\[12pt]
\frac{\mu}{N_Q{+}\mu\bar\mu} {\adag}^Q &
\frac{\mu\bar\mu}{N_Q{+}\mu\bar\mu} \end{pmatrix} \ .
\end{equation}

How do we see that such projectors are BPS? 
Writing $|T\>=\hat{T}|\Hcal\>$, the BPS condition
\begin{equation} \label{TBPS}
a\,|T_i^\ell\> \= |T_i^{\ell'}\>\,\gamma_{\ell'}^{\ \ell}
\qquad\text{implies}\qquad
a\,\hat{T}_i \= \hat{T}_i \,\hat\gamma \qquad\textrm{with}\quad i=1,2
\end{equation}
for some operator $\hat\gamma$ in $\Hcal$. 
We do not know the general solution for arbitrary~$\hat\gamma$.     
However, an important class of solutions arises for the choice $\hat\gamma=a$
where the BPS condition reduces to the ``holomorphicity condition''\footnote{
This may even be the general case: If there exists an invertible operator
$\hat\Gamma$ solving $a\hat\Gamma=\hat\Gamma\hat\gamma$, then the general 
solution to~(\ref{TBPS}) reads $\hat{T}_i=\check{T}_i\hat\Gamma$ with 
$[a,\check{T}_i]=0$, and $\hat\Gamma$ can be scaled to unity.}
\begin{equation}
[\,a\,,\,\hat{T}_i\,] \= 0 \qquad\textrm{for}\quad i=1,2\ .
\end{equation}
These equations are satisfied by {\em any\/} set of functions 
$\{\hat{T}_1,\hat{T}_2\}$ of $a$ alone, i.e.~not depending on~$\adag$.
Indeed, for the example of $\hat{T}_{\text{d}}$ in (\ref{Thatdef}), 
we concretely have
\begin{equation}
\hat\gamma\= a\,P_Q \ +\  (\unity_\Hcal{-}P_Q)\,a\,\sqrt{\sfrac{N-Q}{N}} \ ,
\end{equation}
while $\check{T}(\mu)$ in (\ref{Ttilde}) is obviously holomorphic and thus BPS.
Because $\hat{T}(\mu)$ emerges from $\hat{T}_{\text{d}}$ via a BPS-compatible 
unitary transformation it shares the topological charge~$Q$ and the 
energy $E=8\pi Q$ with the latter.

The generalization to arbitrary values of $n$ and $R<n$ is straightforward:
$\hat{T}$ becomes an $n{\times}R$ array of operators $(\hat{T}_i^L)$, 
on which left multiplication by $a$ amounts to right multiplication by 
an $R{\times}R$ array of operators~$\hat\gamma$. For the special choice
$\hat\gamma_{L'}^{\ L}=\delta_{L'}^{\ L}a$, any collection of holomorphic
functions of $a$ serves as a solution for~$\hat{T}_i^L$.
Quite generally, one can show that for polynomial functions $\hat{T}_i^L(a)$
the resulting projector has a finite topological charge~$Q$ given by the 
degree of the highest polynomial and is thus of finite energy
\cite{Lee:2000ey,Lechtenfeld:2001aw,Foda:2002nt}. 
In this formulation it becomes evident that nonabelian solutions have a 
smooth commutative limit, where $\sqrt{2\theta}a\to{z}$ and $\theta\to0$.
Indeed, they are seen as deformations of the well known solitons in the
$\text{Gr}(n,R)=\frac{\text{U}(n)}{\text{U}(R){\times}\text{U}(n{-}R)}$
Grassmannian sigma model.\footnote{
For the above-discussed example one has $\text{Gr}(2,1)=\C P^1$.}
Hence, the moduli space of the nonabelian solitons coincides with that
of their commutative cousins. By rescaling $\hat{T}\to\hat{T}\Gamma$
with a $\C$-valued $R{\times}R$ matrix~$\Gamma$ we can eliminate $R$
complex parameters from $nR$ independent polynomials. 
For a charge-$Q$ solution, there remain $nRQ+(n{-}1)R$ complex moduli, 
of which $(n{-}1)R$ parametrize the vacuum and $nRQ$ describe the position
and shape of the multi-soliton~\cite{Lee:2000ey}. 
For the case of $\C P^1$ this yields a complex $2Q$ dimensional soliton
moduli space represented by 
$\bigl(\begin{smallmatrix} \check{T}_1 \\ \check{T}_2 \end{smallmatrix}\bigr)
=\bigl(\begin{smallmatrix} \ a^Q+\ldots+\nu \\ 
                     \lambda a^Q+\ldots+\mu \end{smallmatrix}\bigr)$.
Our sample calculation above suggests that taking $\check{T}_2\to\mu$ and
then performing the limit $\mu\to 0$ one recovers the complex $Q$ dimensional 
moduli space of the abelian solitons given by
$|T\>=\bigl(|\a_1\>\,\dots\,|\a_{Q-1}\>\bigr)$ as a boundary.

\subsection{Some non-BPS solutions}
\noindent
For the record we also present a particular class of non-BPS (and non-anti-BPS)
Grassmannian solutions to the equation of motion $\ [\Delta P,P]=0$.
{}From the action of $\Delta$ on a basis operator,
\begin{equation}
\Delta\,|m\>\<n| \= (m{+}n{+}1)\,|m\>\<n| -
\sqrt{mn}\,|m{-}1\>\<n{-}1| - \sqrt{(m{+}1)(n{+}1)}\,|m{+}1\>\<n{+}1| \ ,
\end{equation}
we infer that $\Delta$ maps the $k$-th off-diagonal into itself,
so in particular it retains the diagonal: 
\begin{equation} \label{Deltadiag}
\Delta\,|m\>\<m| \= (2m{+}1)\,|m\>\<m| - 
m\,|m{-}1\>\<m{-}1| - (m{+}1)\,|m{+}1\>\<m{+}1| \ .
\end{equation}
It follows that {\em every\/} diagonal projector is a solution.
Let us first consider the case of U(1).
Given the natural ordering of the basis $\{|m\>\}$ of~$\Hcal$,
any diagonal projector~$P'$ of finite rank~$r{=}Q$ can be written as
\begin{equation} \label{nonBPSP}
P' \= \sum_{s=1}^q \sum_{k=0}^{r_s-1} |m_s{+}k\>\<m_s{+}k|
\= \sum_{s=1}^q S^\+_{m_s}\,P_{r_s}\;S_{m_s}
\qquad\text{with}\quad m_{s+1}>m_s{+}r_s \quad\forall s
\end{equation}
and $\sum_{s=1}^q r_s =r$. Via
\begin{equation} \label{DeltanonBPSP}
\begin{aligned}
\Delta P' &\= \sum_{s=1}^q \Bigl\{
m_s\,|m_s\>\<m_s|\ -\ m_s\,|m_s{-}1\>\<m_s{-}1| \\ &\qquad -\
(m_s{+}r_s)\,|m_s{+}r_s\>\<m_s{+}r_s|\ +\ 
(m_s{+}r_s)\,|m_s{+}r_s{-}1\>\<m_s{+}r_s{-}1| \Bigr\}
\end{aligned}
\end{equation}
its energy is easily calculated as
\begin{equation} \label{nonBPSenergy}
\sfrac{1}{8\pi}\,E[P'] \= Q\ +\ 2\sum_{s=1}^q m_s \ ,
\end{equation}
which is obviously minimized for the BPS case $q=1$ and $m_1=0$.
The energy is additive as long as the two projectors to be combined
are not getting ``too close''.
This picture generalizes to the nonabelian case by formally allowing
$r_s$ and $m_s$ to become infinite.
In particular, the energy does not change when one imbeds $P'$ 
(or several copies of it) into U($\Hcal^{\oplus n}$)
and adds to it a constant projector as in (\ref{specialP}).
Hence, with the proper redefinition of the $m_s$, (\ref{nonBPSenergy}) holds 
for nonabelian diagonal solutions as well.

The inversion $P\to\unity{-}P$ generates additional solutions,
which for the structure in (\ref{specialP}) and BPS-compatible unitaries~$U$
are represented as
\begin{equation} 
P \= U\,\bigl( \widehat{P}-P' \bigr) \, U^\+ \qquad\text{with}\qquad
\widehat{P} = (\unity_n{-}\bar{P})\otimes\unity_\Hcal \qquad\text{and}\qquad
(\unity{-}\widehat{P})P' =0= P'(\unity{-}\widehat{P}) \ ,
\end{equation}
When $\unity{-}\widehat{P}{+}P'$ is BPS, then $P$ becomes anti-BPS 
with topological charge $Q[P]=-\Tr P'$,
producing an anti-soliton with energy $E[P]=8\pi\Tr P'$.
It is possible to combine solitons and anti-solitons to a non-BPS solution via
\begin{equation}
P \= P^{\phantom{\+}}_{\text{sol}} + P_{\overline{\text{sol}}}
\qquad\text{provided}\qquad
P^{\phantom{\+}}_{\text{sol}} P_{\overline{\text{sol}}} \=0\= 
P_{\overline{\text{sol}}} P^{\phantom{\+}}_{\text{sol}}\ ,
\end{equation}
so that their topological charges and energies simply add to
\begin{equation}
Q[P] \= Q[P^{\phantom{\+}}_{\text{sol}}] + Q[P_{\overline{\text{sol}}}] 
\qquad\text{and}\qquad
E[P] \= E[P^{\phantom{\+}}_{\text{sol}}] + E[P_{\overline{\text{sol}}}] \ .
\end{equation}
For the diagonal case, this is included in the solutions discussed above.
Examples for U(1) and U(2) (which appeared in~\cite{Furuta:2002nv}) are
(for $m{>}r$)
\begin{equation}
\begin{aligned}
P &\= \unity_\Hcal - P_m + P_r \= \unity_\Hcal - \sum_{k=r}^{m-1} |k\>\<k|
\qquad&\text{with}&\quad 
Q= r{-}m \ &\text{and}&\quad \sfrac{1}{8\pi}E= r{+}m \ , \\
P &\= P_{r_1} \oplus (\unity_\Hcal{-}P_{r_2}) 
\qquad&\text{with}&\quad
Q= r_1{-}r_2 \ &\text{and}&\quad \sfrac{1}{8\pi}E= r_1{+}r_2 \ ,
\end{aligned}
\end{equation}
respectively. 
Besides the global translations and rotations, other unitary transformations
are conceivably compatible with the equation of motion $\ [\Delta P,P]=0$,
generating moduli spaces of non-BPS Grassmannian solutions. Outside the
Grassmannian manifolds, many more classical configurations are to be found.

\section{Fluctuation analysis}
\noindent
In order to investigate the stability of the classical configurations
constructed in the previous section, we must study the energy functional~$E$
in the neighborhood of the solution under consideration. Since the latter
is a minimum or a saddle point of~$E$, all the linear stability information 
is provided by the Hessian, i.e.~the second variation of~$E$ evaluated at 
the solution. The Hessian is viewed as a linear map on the solution's tangent
space of fluctuations, and its spectrum encodes the invariant information: 
Zero modes belong to field directions of marginal stability
(and extend to moduli if they remain zero to higher orders) while negative
eigenvalues signal instabilities. A perturbation in such a field direction 
provides (part of) the initial conditions of a runaway solution in a time
extension of the model. We cannot be very specific about the stability of
a general classical configuration. Therefore, we shall restrict our attention
to the stability of BPS solutions (as introduced above), at which the 
Hessian simplifies sufficiently to obtain concrete results.
Since any BPS configuration is part of a moduli space which is imbedded
in some Grassmannian which itself lies inside the full configuration space
of the noncommutative U($n$) sigma model, the total fluctuation space contains
the subspace of Grassmannian fluctuations which in turn includes the subspace
of BPS perturbations, the latter being zero modes associated with moduli.
In order to simplify the problem of diagonalizing the Hessian we shall search
for decompositions of the fluctuation space into subspaces which are invariant
under the action of the Hessian. As we consider configurations of finite
energy only, admissible fluctuations~$\phi$ must render $\delta^2 E$ finite 
and keep the ``background'' $\Phi$ unitary. Furthermore, they need to be 
subject to the same conditions as $\Phi$ itself: $\phi$~is bounded, $[a,\phi]$ 
and $[\adag,\phi]$ are Hilbert-Schmidt, and $\Delta\phi$ is traceclass. 
Finally, in keeping with our restricted notion of Grassmannian, we do not 
admit hermitian fluctuations\footnote{
Perturbations inside the Grassmannian must be hermitian 
(see Section~3.2 below).}
which alter the topological charge.

\subsection{The Hessian}
\noindent
The Taylor expansion of the energy functional 
around some finite-energy configuration~$\Phi$ reads
\begin{equation}
\begin{aligned}
E[\Phi{+}\phi] &\= E[\Phi]\,+ 
\int\!\!\text{d}^2z\;\frac{\delta E}{\delta\Phi(z)}[\Phi]\;\phi(z)\,+\,
\frac12 \int\!\!\text{d}^2z\!\int\!\!\text{d}^2z'\;\phi(z)\,
\frac{\delta^2 E}{\delta\Phi(z)\,\delta\Phi(z')}[\Phi]\;\phi(z')\,+ \dots 
\\[4pt]
& \ =:\ E[\Phi]\,+\, E^{(1)}[\Phi,\phi]\,+\, E^{(2)}[\Phi,\phi]\,+\,\dots\ ,
\end{aligned}
\end{equation}
where the U($n$) traces are included in $\int\!\text{d}^2z$.
The perturbation~$\phi$ is to be constrained as to keep the background
$\Phi$ unitary. Since we compute to second order in~$\phi$,
it does not suffice to take $\phi\in T_\Phi\text{U}(\Hcal^{\oplus n})$.
Rather, we must include the leading correction stemming from the exponential
map onto $\text{U}(\Hcal^{\oplus n})$, which generates the finite 
perturbation $\phi=\Phi'{-}\Phi$. The latter is subject to the constraint
\begin{equation} \label{phidagger}
(\Phi^\+{+}\phi^\+)(\Phi{+}\phi) \= \unity \qquad\Longrightarrow\qquad \phi^\+
\= -\Phi^\+\phi\,\Phi^\+ + \Phi^\+\phi\,\Phi^\+\phi\,\Phi^\+ + O(\phi^3)\ ,
\end{equation}
which we shall use to eliminate $\phi^\+$ from the variations.
It is important to realize that in this way the term linear in $\phi^\+$ 
generates a contribution to $E^{(2)}[\Phi,\phi]$.
Performing the expansion for the concrete expression (\ref{E}) and 
using (\ref{phidagger}) we arrive at\footnote{
The same result is obtained by considering $E[\exp\{t\phi\}\Phi]$ up
to $O(t^2)$.}
\goodbreak
\begin{align}
E^{(1)}[\Phi,\phi] &\= \pi\,\Tr\bigl\{ 
[\adag,\,\Phi^\+\phi\,\Phi^\+]\,[a\,,\,\Phi] + 
[\adag,\,\Phi]\,[a\,,\,\Phi^\+\phi\,\Phi^\+] -
[\adag,\,\Phi^\+]\,[a\,,\,\phi] -
[\adag,\,\phi]\,[a\,,\,\Phi^\+] 
\bigr\} \nonumber\\[4pt]
&\= 2\pi\,\Tr\bigl\{ 
(\Delta\Phi^\+\,\Phi - \Phi^\+\,\Delta\Phi)\,\Phi^\+\,\phi\bigr\} \= 0\ , 
\\[6pt]
E^{(2)}[\Phi,\phi] &\= \pi\,\Tr\bigl\{
[\adag,\,\Phi^\+\phi\,\Phi^\+]\,[a\,,\,\phi] +
[\adag,\,\phi]\,[a\,,\,\Phi^\+\phi\,\Phi^\+] \nonumber\\  &\qquad\ \ -\
[\adag,\,\Phi^\+\phi\,\Phi^\+\phi\,\Phi^\+]\,[a\,,\,\Phi] -
[\adag,\,\Phi]\,[a\,,\,\Phi^\+\phi\,\Phi^\+\phi\,\Phi^\+] 
\bigr\} \nonumber\\[4pt]
&\= 2\pi\,\Tr\bigl\{
\Phi^\+\phi\,\Phi^\+\phi\,\Phi^\+\,\Delta\Phi - 
\Phi^\+\phi\,\Phi^\+\,\Delta\phi \bigr\} \nonumber\\[4pt]
&\= 2\pi\,\Tr\bigl\{
\phi^\+\,\Delta\phi - \phi^\+\,(\Phi\Delta\Phi^\+)\;\phi \} + O(\phi^3) \ =:\
2\pi\,\Tr\bigl\{ \phi^\+\,H\,\phi \bigr\} + O(\phi^3) \ ,
\label{Qform}
\end{align}
defining the Hessian $H=\Delta-(\Phi\Delta\Phi^\+)$ as a self-adjoint operator.
Hence, our task is essentially reduced to working out the spectrum of the 
Hessian.  Since $\Delta$ is clearly a positive semidefinite operator,
an instability can only occur in directions for which $\<\Phi\Delta\Phi^\+\>$
is sufficiently large.

For later reference, we present the action of $H$ in the oscillator basis,
\begin{equation} \label{Hosc}
\begin{aligned}
{}& H \ \sum_{m,\ell} \phi_{m,\ell}\,|m\>\<\ell| \= 
\sum_{m,\ell} (H\phi)_{m,\ell}\,|m\>\<\ell|
\qquad\qquad\text{with} \\[4pt]
{}& (H\phi)_{m,\ell} \= (m{+}\ell{+}1)\,\phi_{m,\ell} -
\sqrt{(m{+}1)(\ell{+}1)}\,\phi_{m+1,\ell+1} - 
\sqrt{m\ell}\,\phi_{m-1,\ell-1} \\[4pt] 
{}& \qquad\qquad -\ \sum_{j,k} \Phi_{m,j}\,\bigl\{ 
(j{+}k{+}1)\,\Phi_{j,k} -
\sqrt{(j{+}1)(k{+}1)}\,\Phi_{j+1,k+1} - 
\sqrt{jk}\,\Phi_{j-1,k-1} 
\bigr\}\,\phi_{k,\ell} \ ,
\end{aligned}
\end{equation}
where $\Phi_{m,\ell}$ as well as $\phi_{m,\ell}$ are still $n{\times}n$
matrix-valued. At diagonal abelian backgrounds, as given in (\ref{nonBPSP}),
the matrices reduce to numbers and the latter expression simplifies to
\begin{equation} \label{Hdiag}
\begin{aligned}
(H\phi)_{m,\ell} &\= (m{+}\ell{+}1-2b_m)\,\phi_{m,\ell} -
\sqrt{(m{+}1)(\ell{+}1)}\,\phi_{m+1,\ell+1} - 
\sqrt{m\ell}\,\phi_{m-1,\ell-1}
\qquad\text{with} \\[4pt] 
b_m &\= \sum_{s=1}^q \bigl\{ 
\delta_{m,m_s}m_s + \delta_{m,m_s-1}m_s +
\delta_{m,m_s+r_s}(m_s{+}r_s) + 
\delta_{m,m_s+r_s-1}(m_s{+}r_s) \bigr\} \ ,
\end{aligned}
\end{equation}
which for fixed $\ell$ differs from $(\Delta\phi)_{m,\ell}$ in at most
$4q$ entries.  For diagonal U(1) BPS backgrounds $\Phi=\Phi_r=\unity-2P_r$
this reduces further to
\begin{equation} \label{bdiag}
b_m \= r\,\bigl( \delta_{m,r-1} + \delta_{m,r} \bigr) \ .
\end{equation}
It is important to note that these expressions do not yet yield a
matrix representation of $H$ because the constraint~(\ref{phidagger})
on the allowable perturbations must still be taken into account. 
We can do this by replacing $\phi$ with $\phi-\Phi\phi^\+\Phi$
everywhere; then $\phi$ becomes unconstrained.

\subsection{Decomposition into even and odd fluctuations}
\noindent
We specialize to Grassmannian backgrounds, $\Phi=\unity-2P=\Phi^\+$,
characterized by a hermitian projector~$P$ and obeying $\Phi^2=\unity$. 
Any such projector induces an orthogonal decomposition
\begin{equation}
\C^n\otimes\Hcal\= P\,(\C^n\otimes\Hcal)\oplus(\unity{-}P)(\C^n\otimes\Hcal)
\ =:\ \text{im}P \oplus \text{ker}P \ ,
\end{equation}
and a fluctuation~$\phi$ decomposes accordingly as
\begin{equation}
\phi \= 
\underbrace{P\,\phi\,P\ +\ (\unity{-}P)\,\phi\,(\unity{-}P)}_{\phi_{\text{e}}}
\ +\
\underbrace{P\,\phi\,(\unity{-}P)\ +\ (\unity{-}P)\,\phi\,P}_{\phi_{\text{o}}} 
\ ,
\end{equation}
where the subscripts refer to ``even'' and ``odd'', respectively.
Since $\Phi$ acts as $-\unity$ on $\text{im}P$ but as $+\unity$ on
$\text{ker}P$, we infer that
\begin{equation} \label{hermfluct}
\Phi\,\phi_{\text{e}} \= \phi_{\text{e}} \Phi \qquad\text{and}\qquad
\Phi\,\phi_{\text{o}} \=-\phi_{\text{o}} \Phi \qquad\Longrightarrow\qquad
\phi^\+_{\text{e}} \=-\phi_{\text{e}} \qquad\text{and}\qquad
\phi^\+_{\text{o}} \= \phi_{\text{o}}
\end{equation}
to leading order from (\ref{phidagger}),
i.e.~even fluctuations are anti-hermitian while odd ones are hermitian.
This implies that odd fluctuations keep $\Phi$ inside its Grassmannian,
but even ones perturb away from it.
It also follows that in
\begin{equation} \label{cross1}
\Tr \bigl( \phi^\+\,\Delta\phi \bigr) \=
\Tr \bigl( \phi^\+_{\text{e}}\,\Delta\phi_{\text{e}} \bigr) +
\Tr \bigl( \phi^\+_{\text{o}}\,\Delta\phi_{\text{o}} \bigr) +
\Tr \bigl( \phi^\+_{\text{e}}\,\Delta\phi_{\text{o}} \bigr) +
\Tr \bigl( \phi^\+_{\text{o}}\,\Delta\phi_{\text{e}} \bigr)
\end{equation}
the last two terms cancel each other. Furthermore, 
the equation of motion (\ref{Peom}), $[\Delta P,P]=0$, implies that
\begin{equation} \label{cross2}
(\Phi\Delta\Phi)_{\text{o}} \= 0 \qquad\Longrightarrow\qquad
\Tr \bigl( \phi^\+_{\text{e}}\;\Phi\Delta\Phi\;\phi_{\text{o}} \bigr) \= 0 \=
\Tr \bigl( \phi^\+_{\text{o}}\;\Phi\Delta\Phi\;\phi_{\text{e}} \bigr) \ ,
\end{equation}
because only an even number of odd terms in a product survives under the trace.
Combining (\ref{cross1}) and (\ref{cross2}) we conclude that
\begin{equation} \label{eosplit}
E^{(2)}[\Phi,\phi_{\text{e}}{+}\phi_{\text{o}}] \=
E^{(2)}[\Phi,\phi_{\text{e}}] + E^{(2)}[\Phi,\phi_{\text{o}}] \ ,
\end{equation}
which allows us to treat these two types of fluctuations separately.

The above decomposition has another perspective.
Recall that any background configuration $\Phi$ being unitary
can be diagonalized by some unitary transformation,
\begin{equation} \label{diagPhi}
\Phi \= U\,\Phi_{\text{d}}\,U^\+
\= U\,\text{diag}\bigl(\{\e^{\ic\lambda_i}\}\bigr)\,U^\+\ .
\end{equation}
When $\Phi$ is hermitian, i.e.~inside some Grassmannian, the diagonal phase
factors can be just $+1$ or $-1$, and $U$ is determined only up to a factor
$V\in\text{U(im}P)\times\text{U(ker}P)$
which keeps the two eigenspaces $\text{im}P$ and $\text{ker}P$ invariant. 
Adding a perturbation~$\phi$ lifts the high degeneracy of $\Phi$,
so that the diagonalization of $\Phi{+}\phi$ requires an infinitesimal
``rotation'' $K$ of $\text{im}P$ and $\text{ker}P$ inside~$\Hcal$ as well as 
a ``large'' rediagonalization $V$ inside the two eigenspaces. 
Modulo higher order terms we may write
\begin{equation}
\begin{aligned}
\Phi + \phi &\= U\,(1{+}K)\,V\,
(\Phi_{\text{d}}+\phi_{\text{d}})\,V^\+(1{-}K)\,U^\+ \= \Phi\ +\ U \bigl\{ 
V\,\phi_{\text{d}}\,V^\+\,+\,[K\,,\,\Phi_{\text{d}}] \bigr\} U^\+ \\[4pt]
& \text{with}\qquad [V,\Phi_{\text{d}}]=0 \qquad\text{and}\qquad
K = -K^\+ \quad\text{infinitesimal}\ ,
\end{aligned}
\end{equation}
where $\phi_{\text{d}}$ is a purely diagonal and anti-hermitian fluctuation. 
Since $V$ depends on $\phi$ it should not be absorbed into~$U$.
It rather generates all non-diagonal fluctuations inside 
$\text{im}P$ and $\text{ker}P$, allowing us to rewrite
\begin{equation}
V\,\phi_{\text{d}}\,V^\+ \= \phi'_{\text{d}}\ +\ 
[ \Lambda_{\text{e}}\,,\,\phi_{\text{d}} ]\ ,
\qquad\text{with}\qquad \Lambda_{\text{e}}^\+ = -\Lambda_{\text{e}}
\end{equation}
being a generator of $\text{U(im}P)\times\text{U(ker}P)$ and
a modified diagonal perturbation~$\phi'_{\text{d}}$.
Redenoting also $K=\epsilon\Lambda_{\text{o}}$ with a real and
infinitesimal $\epsilon$ and a generator $\Lambda_{\text{o}}$ of the
Grassmannian, the general fluctuation is parametrized as
\begin{equation}
\phi \= U\,\bigl\{ \phi'_{\text{d}}\ +\ 
[ \Lambda_{\text{e}}\,,\,\phi_{\text{d}} ]\ +\
[ \Lambda_{\text{o}}\,,\,\epsilon\Phi_{\text{d}} ] \bigr\}\, U^\+
\end{equation}
and decomposed (after diagonalizing the background via $U$) into
a ``radial'' part $\phi'_{\text{d}}$ and an ``angular'' part
$\phi_{\text{a}}=[\Lambda,\text{any}]$ with $\Lambda$ generating 
U($\Hcal^{\oplus n}$)~\cite{Gopakumar:2000zd}. 
For a Grassmannian background all terms have definite hermiticity properties,
and we can identify
\begin{equation}
\phi_{\text{e}} \= U\,\bigl\{ \phi'_{\text{d}}\ +\
[ \Lambda_{\text{e}}\,,\,\phi_{\text{d}} ] \bigr\}\, U^\+
\qquad\text{and}\qquad
\phi_{\text{o}} \= 
U\,\bigl\{ [ \Lambda_{\text{o}}\,,\,\epsilon\Phi_{\text{d}} ] \bigr\}\, U^\+ 
\= \epsilon\,[ U \Lambda_{\text{o}} U^\+\,,\Phi ] \ .
\end{equation}
We have seen in (\ref{eosplit}) above that the even and odd fluctuations can 
be disentangled in~$E^{(2)}$. It is not clear, however, whether the diagonal
perturbations can in turn be separated from the even angular ones in the
fluctuation analysis (but see below for diagonal backgrounds where $U=\unity$).

\subsection{Odd or Grassmannian perturbations}
\noindent
As far as stability of BPS configurations is concerned, the odd
perturbations are easily dealt with by a general argument.
Since a shift by $\phi_{\text{o}}$ keeps $\Phi$ inside its Grassmannian,
wherein $\Phi$ already minimizes the energy, such a perturbation cannot lower
the energy any further and we can be sure that negative modes are absent here.
Therefore, solitons in the noncommutative {\em Grassmannian\/} sigma model
are stable, up to possible zero modes.
An obvious zero mode is generated by the translational and rotational symmetry.
A glance at (\ref{globaltrans}) and (\ref{globalrot}) shows that
the corresponding infinitesimal operators $\Lambda$ are given by
\begin{equation}
\Lambda_{\text{trans}} \= \a\,\adag - \ab\,a 
\qquad\text{and}\qquad
\Lambda_{\text{rot}} \= \ic\vartheta\,\adag a\ ,
\end{equation}
which indeed leads to the annihilation of $[\Lambda_{\text{trans}},\Phi]$
and $[\Lambda_{\text{rot}},\Phi]$ by~$H$ as we shall see. 

{}From the discussion at the end of Section~2.4
we know that for $r{>}1$ there are additional zero modes inside the
Grassmannian, because the multi-soliton moduli spaces are higher-dimensional.
Parametrizing these BPS-compatible perturbation as follows,
\begin{equation} \label{BPSpert}
\Phi + \phi_{\text{o}} \= U^{\phantom{\+}}_{\text{B}}\,\Phi\, U^\+_{\text{B}}
\= \Phi + \epsilon\,[\Lambda^{\phantom{\+}}_{\text{B}},\Phi] + O(\epsilon^2)
\qquad\text{for}\quad 
U^{\phantom{\+}}_{\text{B}} = \e^{\epsilon\Lambda^{\phantom{\+}}_{\text{B}}}
\quad\text{with}\quad 
\Lambda^\+_{\text{B}} = -\Lambda^{\phantom{\+}}_{\text{B}} \ ,
\end{equation}
the corresponding Lie-algebra element (\ref{defomega}) becomes
\begin{equation}
\omega \= U^\+_{\text{B}}\,[a\,,U^{\phantom{\+}}_{\text{B}}] 
\= (\e^{-\epsilon\,\text{ad}\Lambda^{\phantom{\+}}_{\text{B}}}-1)\,a 
\= \epsilon\,[a\,,\Lambda^{\phantom{\+}}_{\text{B}}] + O(\epsilon^2)\ ,
\end{equation}
and the condition~(\ref{BPScomp}) of BPS compatibility 
to leading order in $\epsilon$ reads
\begin{equation} \label{BPSfluct}
[a\,,\Lambda^{\phantom{\+}}_{\text{B}}]\,|T\> \= |T\>\,\gamma_\omega
\qquad\text{for some $r{\times}r$ matrix $\gamma_\omega$}\ .
\end{equation}

Let us try to find $\Lambda^{\phantom{\+}}_{\text{B}}$ in the abelian case 
by perturbing around the diagonal BPS configuration
\begin{equation}
|T_r\> \= \bigl( |0\>\ |1\>\ \dots\ |r{-}1\> \bigr)
\qquad\Longleftrightarrow\qquad
\Phi_r \= \unity_\Hcal\ -\ 2\sum_{m=0}^{r-1} |m\>\<m| \ .
\end{equation}
Expanding the generator
\begin{equation}
\Lambda^{\phantom{\+}}_{\text{B}}\=\sum_{m,\ell}\Lambda_{m,\ell}\,|m\>\<\ell|
\qquad\text{with}\quad \bar\Lambda_{m,\ell} = -\Lambda_{\ell,m}
\end{equation}
it is easy to see that the even part of $\Lambda^{\phantom{\+}}_{\text{B}}$
automatically fulfils~(\ref{BPSfluct}), and so restrictions to 
$\Lambda_{m,\ell}$ arise only for the odd components.
In fact, only the terms with $m{\ge}r$ and $\ell{<}r$ in the above sum
may violate~(\ref{BPSfluct}), which leads to the conditions
\begin{equation} \label{Leqs}
\sqrt{m{+}1}\,\Lambda_{m+1,\ell} - \sqrt{\ell}\,\Lambda_{m,\ell-1} \= 0
\qquad\text{for all}\quad m\ge r \quad\text{and}\quad \ell<r \ . 
\end{equation}
Since $\ell{=}0$ yields $\Lambda_{m+1,0}{=}0$ as a boundary condition, 
this hierarchy of equations puts most components to zero, except for 
$m=r,\dots,r{+}\ell$ at any fixed $\ell{<}r$. These remaining $r(r{+}1)/2$ 
components are subject to $r(r{-}1)/2$ equations from~(\ref{Leqs}), 
whose solution
\begin{equation} \label{Lsol}
\Lambda_{r+j,\ell+j} \= \sqrt{ \sfrac{
(\ell{+}j)(\ell{+}j{-}1)\cdots(\ell{+}1)}{(r{+}j)(r{+}j{-}1)\cdots(r{+}1)}}\,
\Lambda_{r,\ell} 
\qquad\text{for}\quad j=1,\dots,r{-}1{-}\ell 
\quad\text{at}\quad \ell\le r{-}2
\end{equation}
fixes $r(r{-}1)/2$ components in terms of the $r{-}1$ components
appearing on the right hand side, which therefore are free complex parameters.
The $r$th free parameter $\Lambda_{r,r-1}$ does not enter and is associated
with the rigid translation mode.\footnote{
The rigid rotation mode is absent because $\Phi_r$ is spherically symmetric.}
Finally, the ensueing BPS perturbation~(\ref{BPSpert}) is found to be
\begin{equation} \label{BPSmode}
\phi_{\text{o}} \= \epsilon\,[\Lambda^{\phantom{\+}}_{\text{B}},\Phi_r] \=
\epsilon \sum_{m=r}^{\infty} \sum_{\ell=0}^{r-1} \Bigl(
\Lambda_{m,\ell}\,|m\>\<\ell|\ +\ \bar\Lambda_{m,\ell}\,|\ell\>\<m| \Bigr) \ ,
\end{equation}
with $\Lambda_{m,\ell}$ taken from (\ref{Lsol}).
To higher orders in~$\epsilon$, the BPS-compatibility condition is not
automatically satisfied by our solution~(\ref{Lsol}) but this can be repaired
by adding suitable even components to~$\Lambda^{\phantom{\+}}_{\text{B}}$.
Our result ties in nicely with the observation of $r$ complex moduli~$\a_k$ in
(\ref{Pcoherentr}) whose shifts produce precisely $r$ complex zero modes.
For the simplest non-trivial case of $r{=}2$, one can also extract these modes
from differentiating (\ref{U2}) with respect to $\a_1$ or~$\a_2$.

\subsection{Even or non-Grassmannian perturbations}
\noindent
The stability analysis of BPS configurations inside the full noncommutative
U($n$) sigma model requires the investigation of the even fluctuations
$\phi_{\text{e}}$ as well. Here, we have only partial results
to offer.\footnote{
Even for the commutative sigma model this is an open problem 
\cite{Zakrzewski}.}
Yet, there is the following general argument which produces an unstable
even fluctuation mode~$\phi_{\text{neg}}$ (but not an eigenmode) 
for {\em any\/} noncommutative multi-soliton $\Phi=\unity-2P$ with $Q>0$.
For this, consider some other multi-soliton $\tilde\Phi=\unity-2\tilde P$
which is contained in $\Phi$ in the sense that
\begin{equation}
\text{im}(\tilde P) \ \subset\ \text{im}(P)
\qquad\Longleftrightarrow\qquad
\tilde P\,P \= P\,\tilde P \= \tilde P \ .
\end{equation}
We then simply say that $\tilde P\subset P$.
It follows that their difference $\Pi$ is the orthogonal complement
of $\tilde P$ in im($P$),
\begin{equation} \label{diffP}
\Pi \= P - \tilde P \ \subset\ P \qquad\Longrightarrow\qquad
\Pi^2 = \Pi \qquad\text{and}\qquad \Pi\,\tilde P =0= \tilde P\,\Pi \ .
\end{equation}
In particular, we may choose $\tilde P=0$.
For any such pair ($P,\tilde P$) there exists a continuous path 
\begin{equation} \label{path}
\begin{aligned}
\Phi(s) \= \e^{\ic s\Pi} (\unity{-}2P) \=
\unity-2P +(1{-}\e^{\ic s})\Pi\=\unity-2\tilde P -(1{+}\e^{\ic s})\Pi \\[4pt]
\text{connecting} \qquad \Phi(0)=\Phi=\unity{-}2P \qquad
\text{with} \qquad \Phi(\pi)=\tilde\Phi=\unity{-}2\tilde P \ .
\end{aligned}
\end{equation}
Note that $\Phi(s)$ interpolates between two different Grassmannians,
touching them only at $s{=}0$ and $s{=}\pi$.
Since we assumed that $\Phi$ and $\tilde\Phi$ are BPS we know that
\begin{equation}
E[\Phi] \= 8\pi\,Q  \qquad\text{and}\qquad E[\tilde\Phi] \= 8\pi\,\tilde Q
\end{equation}
with $Q$ and $\tilde Q$ being the topological charges of $P$ and $\tilde P$,
respectively.\footnote{
Note that $\tilde Q$ need not be smaller than $Q$ when $r(P)$ is infinite.}
Inserting (\ref{path}) into the expression (\ref{E}) for the energy
and abbreviating $1{-}\e^{\ic s}=\rho$ we compute
\begin{align}
E(s) \ :=\ E[\Phi(s)]
&\= 8\pi\,Q
-2\pi(\rho{+}\bar\rho)\,\Tr\bigl([a,P][\Pi,\adag]+[\adag,P][\Pi,a]\bigr)
+2\pi\,\rho\bar\rho\,\Tr\bigl([a,\Pi][\Pi,\adag]\bigr) \nonumber \\[4pt]
&\= 8\pi\,Q -4\pi(1{-}\cos s)\,
\Tr\bigl([a,P][\Pi,\adag]+[\adag,P][\Pi,a]-[a,\Pi][\Pi,\adag]\bigr)\ ,
\label{Escalc}
\end{align}
where the two traces could be combined due to the relation
\begin{equation}
\rho + \bar\rho \= \rho\bar\rho \= 2(1{-}\cos s) \ .
\end{equation}
Luckily, we do not need to evaluate the traces above.
Knowing that $E(\pi)=8\pi\tilde Q$ we infer that the last trace in
(\ref{Escalc}) must be equal to $Q{-}\tilde Q$
and hence
\begin{equation} \label{Epath}
\begin{aligned}
E(s) &\= 8\pi\,Q\,-\,4\pi(1{-}\cos s)(Q-\tilde Q)
\=8\pi\bigl(Q\,\cos^2\sfrac{s}{2}\,+\,\tilde Q\,\sin^2\sfrac{s}{2}\bigr)\\[4pt]
&\= 8\pi\,Q\,-\,2\pi(Q-\tilde Q)\,s^2\,+\,O(s^4)\ .
\end{aligned}
\end{equation}
Evidently,
for $Q>0$ we can always lower the energy of a given BPS configuration
by applying an even perturbation $\phi_{\text{neg}}=-\ic\epsilon\Pi$ towards a
BPS solution with smaller charge $\tilde Q<Q$.
The higher the charge of $P$ the more such modes are present.
Yet, they are not independent of one another but rather span a cone
extending from the background, as we shall see in examples below. In fact,
there is no reason to expect any of these unstable fluctuations to represent
an eigenmode of the Hessian, and in general they do not. Nevertheless, their
occurrence again demonstrates that there must be (at least) one negative
eigenvalue of~$H$, and for diagonal U(1) BPS backgrounds we shall prove in 
Section~4.3 that there is exactly one.
The only stable BPS solutions are therefore the ``vacua'' defined by
$P'{=}0$ in (\ref{specialP}) and based on a constant U($n$)
projector~$\bar{P}$. This leaves no stable solutions in the abelian case
besides $\Phi=\unity_\Hcal$.

In addition to the unstable mode, there exist also a number of non-Grassmannian
zero modes around each BPS configuration, which generate nearby non-BPS
solutions to the equation of motion. This will become explicit in the examples
discussed below.

\section{Perturbations of U(1) backgrounds}
\subsection{Invariant subspaces}
\noindent
We specialize further to diagonal U(1) backgrounds $\Phi$ as given by
(\ref{nonBPSP}) (not necessarily BPS).
Using (\ref{Hosc}) it is easy to see that $H$ maps any off-diagonal 
into itself. Let us parametrize the $k$-th upper diagonal $\Dcal_k$ as
\begin{equation} \label{Dkdef}
\phi_{(+k)} \= \sum_{m=0}^\infty \mu_{(k)m}\,|m\>\<m{+}k| 
\qquad\text{with}\quad \mu_{(k)m} \in\C
\qquad\text{for}\quad k=0,1,2,\dots\ .
\end{equation}
The hermiticity properties~(\ref{hermfluct}) of the perturbations demand that 
we combine $\Dcal_k$ and $\Dcal_k^\+$ into a subspace $\Ecal_k$ of the $k$-th
upper plus lower diagonals by defining
\begin{equation}
\Ecal_k\ :=\ \bigl\{ \phi_{(k)}\,\bigm|\, 
\phi_{(k)} \= \phi_{(+k)} - \Phi\,\phi_{(+k)}^\+ \Phi \bigr\}\ ,
\qquad\text{which implies}\qquad
\phi_{(k)}^\+ \= - \Phi\,\phi_{(k)}\,\Phi \ .
\end{equation}
The direct sum of all $\Ecal_k$ is the full admissible tangent space to 
U($\Hcal$), and so we may decompose
\begin{equation} \label{finesplit}
\phi \= \sum_{k=0}^\infty \phi_{(k)} \= 
\sum_{k=0}^\infty \sum_{m=0}^\infty 
\Bigl\{ \mu_{(k)m}\,|m\>\<m{+}k|\ \mp\ \bar\mu_{(k)m}\,|m{+}k\>\<m| \Bigr\}\ ,
\end{equation}
with the sign depending on whether the component is even or odd. 
Since $H$ maps $\Ecal_k$ into itself, the bilinear form 
defined in (\ref{Qform}) is block-diagonal on the set of $\Ecal_k$,
\begin{equation}
2\pi\,\Tr \bigl\{ \phi_{(k)}^\+\,H\,\phi_{(\ell)} \bigr\} \ 
\sim\ \delta_{k\ell} \ ,
\end{equation}
which implies the factorization
\begin{equation} \label{offdiag}
E^{(2)}\,[\Phi,\textstyle{\sum_k}\phi_{(k)}] \= 
\sum_k E^{(2)}\,[\Phi,\phi_{(k)}] \ .
\end{equation}
In other words,
$\Ecal_k$ forms an $H$-invariant subspace for each value of~$k$.
In particular, $\Ecal_0$ is the space of admissible skew-hermitian diagonal 
matrices, and the (purely imaginary) diagonal fluctuations
$\phi_{(0)}\equiv\phi_{\text{d}}\in\Ecal_0$ can be considered on their own.

\subsection{Results for diagonal U(1) backgrounds}
\noindent
Apparently, for diagonal U(1) backgrounds it suffices to study the second 
variation form $E^{(2)}[\Phi,.]$ on each subspace $\Ecal_k$ ($k\ge0$) 
separately. 
Let us give more explicit formulae for the restriction of $E^{(2)}[\Phi,.]$ 
to these subspaces. To this end, we denote the non-zero entries of $\Phi$ 
by $\de_j:=\Phi_{jj}$ ($j=0,1,2,\dots$). 
We have $\de_j=\pm 1$ for each $j$, and the set
\begin{equation}
J\ :=\ \{j\ge0\;|\;\de_{j+1}\neq\de_j\}
\end{equation}
is finite. Then, a straightforward calculation shows that the restriction 
of $E^{(2)}[\Phi,.]$ to
\begin{equation} \label{phi0}
\begin{aligned}
&\Ecal_0 \= \bigl\{ \phi_{(0)}\,\bigm|\,
\phi_{(0)} \= \ic \sum_{m=0}^\infty \phi_m\,|m\>\<m| 
\quad\text{with}\quad \phi_m \in\R \bigr\}
\qquad\text{is given by} \\
&\sfrac{1}{2\pi} E^{(2)}[\Phi,\phi_{(0)}] \= 
\sum_{m=0}^\infty (m{+}1)\,(\phi_{m+1}-\phi_m)^2
-2 \sum_{j\in J} (j{+}1)\,(\phi_{j+1}^2 + \phi_j^2) \ .
\end{aligned}
\end{equation}

The formula for $E^{(2)}[\Phi,.]$ on $\Ecal_{k>0}$ is more complicated.
Considering a fixed $k$-th upper diagonal $\Dcal_k$ as parametrized in
(\ref{Dkdef}), we first evaluate (suppressing the subscript $(k)$)
\begin{equation}
\Tr \{ \phi^\+_{(+k)}\,\Delta\,\phi_{(+k)} \} \= \sum_{m=0}^\infty
|\sqrt{m{+}1}\,\mu_{m+1}-\sqrt{m{+}k{+}1}\,\mu_m|^2 \=
\sfrac12 k\,|\mu_0|^2 + \sfrac12 \sum_{m=0}^\infty R_m(\mu) \ ,
\end{equation}
with
\begin{align} \label{Rmdef}
R_m(\mu) &\ :=\
|\sqrt{m{+}1}\,\mu_{m+1}-\sqrt{m{+}k{+}1}\,\mu_{m}|^2+
|\sqrt{m{+}k{+}1}\,\mu_{m+1}-\sqrt{m{+}1}\,\mu_{m}|^2 \\[6pt] \nonumber
&\= \bigl(\sqrt{m{+}k{+}1}-\sqrt{m{+}1}\bigr)^2 
\bigl(|\mu_{m+1}|^2+|\mu_m|^2\bigr)+
2\sqrt{m{+}k{+}1}\sqrt{m{+}1}\,|\mu_{m+1}-\mu_m|^2\ .
\end{align}
Armed with this expressions, we compute the second variation on~$\Ecal_k$.
To state the outcome, it is useful to introduce the index sets
\begin{equation}
J-k\ :=\ \{j{-}k\in\NN_0\,|\, j\in J\} \qquad\text{and}\qquad
A\ :=\ (J\cup(J-k))\setminus (J\cap(J-k)) \ ,
\end{equation}
i.e.~$A$ is the symmetric difference of $J$ and $J{-}k$.
For $\phi_{(k)}\in\Ecal_k$ a direct calculation results in
\begin{align} \label{QPhik}
\sfrac{1}{2\pi} E^{(2)}[\Phi,\phi_{(k)}] &\= 
k\,|\mu_0|^2\ +\sum_{j\in\NN_0\setminus A}R_j(\mu)\ +\
\sum_{j\in A}(2j{+}2{+}k)\bigl(|\mu_j|^2+|\mu_{j+1}|^2\bigr) \\[6pt] \nonumber
&\quad -2\,\sum_{j\in J}\,(j{+}1)\bigl(|\mu_j|^2+|\mu_{j+1}|^2\bigr)\ -\
2\!\sum_{j\in J-k}\!(j{+}k{+}1)|\mu_j|^2\ -\
2\!\!\!\sum_{j\in J-k+1}\!\!\!(j{+}k)|\mu_j|^2 \ .
\end{align}

This expression simplifies when $k$ is greater than the largest element
of~$J{+}1$. Then the sets $J{-}k$ and $J{-}k{+}1$ are empty,
so $A=J$, and (\ref{QPhik}) takes the form
\begin{equation} \label{QPhik2}
\sfrac{1}{2\pi} E^{(2)}[\Phi,\phi_{(k)}] \= 
k\,|\mu_0|^2\ +\sum_{j\in\NN_0\setminus J}R_j(\mu)
\ +\ k\,\sum_{j\in J}\bigl(|\mu_j|^2+|\mu_{j+1}|^2\bigr) \ .
\end{equation}
One sees that for each $j{\in}J$ the corresponding coefficient $\mu_j$ 
decouples from all successive coefficients~$\mu_{m>j}$.
Since the elements of~$J$ signify in the string of $\mu_m$ the boundaries
between even and odd fluctuations, this observation confirms the
decomposition~(\ref{eosplit}) of~$E^{(2)}$ into an even and odd part also after
the restriction to~$\Ecal_k$.\footnote{
For $|J|>1$ the even and/or odd part of $E^{(2)}$ is split further. In total,
$E^{(2)}$ decomposes into at least $|J|{+}1$ blocks.}
Furthermore, we note that $R_j(\mu)>0$ unless $\mu_j=\mu_{j+1}=0$,
which makes it obvious from the expression~(\ref{QPhik2}) that the
quadratic form~$E^{(2)}$ is strictly positive on each $\Ecal_k$ with
$k>\max_{j\in J}(j{+}1)$. 
For the remaining~$\Ecal_k$ we have to work a little harder.

\subsection{Results for diagonal U(1) BPS backgrounds}
\noindent
The idea is to pursue the {\it reduction of~$E^{(2)}$ to a sum of squares\/}
whenever possible. For diagonal BPS solutions~(\ref{diagonalBPS})
\begin{equation}
\Phi_r \= \unity_\Hcal\ -\ 2\sum_{m=0}^{r-1} |m\>\<m| 
\= \sum_{m=0}^\infty \de_m\,|m\>\<m|
\qquad\text{with}\quad \de_m \=
\begin{cases} -1 & \text{for $m<r$} \\ +1 & \text{for $m\ge r$} \end{cases}\ ,
\end{equation}
this strategy turns out to be successful at all $k\ge1$ 
but breaks down at $k=0$.
We note that now $J=\{r{-}1\}$, which implies a distinction of cases:
``very off-diagonal'' perturbations have $k>r$, 
``slightly off-diagonal'' perturbations occur for $1\le k\le r$,
and diagonal perturbations mean $k=0$, to be discussed last.
It will also be instructive to visualize the Hessian on the space $\Ecal_k$ 
in matrix form,
\begin{equation}
E^{(2)}[\Phi_r,\phi_{(k)}]\=
2\pi\sum_{m,\ell=0}^\infty \bar\mu_m\,H^{(k)}_{m\ell}\,\mu_\ell\ ,
\end{equation}
defining an infinite-dimensional matrix $H^{(k)}=\bigl(H^{(k)}_{m\ell}\bigr)$.

\subsubsection{Very off-diagonal perturbations}
\noindent
Since $J$ consists just of one element, at $k>r$ the matrix~$H^{(k)}$
splits into two parts only, of which the odd one has finite size~$r$.
More explicitly,
\begin{equation}
H^{(k)} \= H^{(k)}_{\text{Gr}(P)}\,\oplus\,H^{(k)}_{\text{ker}P} \ ,
\end{equation}
with
\begin{align}
\sfrac{1}{2} H^{(k)}_{\text{Gr}(P)} = \begin{pmatrix}
k+1 & -\sqrt{1(k{+}1)} & & & & \\[8pt]
-\sqrt{1(k{+}1)} & k+3 & -\sqrt{2(k{+}2)} & & & \\[8pt]
 & -\sqrt{2(k{+}2)} & k+5 & \qquad\ddots\!\!\!\!\!\!\!\! & & \\[8pt]
 & & \!\!\!\!\!\!\!\!\ddots & \;\ddots & \ddots & \\[8pt]
 & & & \!\!\!\!\!\!\!\!\!\!\!\!\!\!\!\!\ddots 
                       & 2r+k-3 & -\sqrt{(r{-}1)(r{+}k{-}1)} \\[8pt]
 & & & & \!\!\!\!\!\!\!\!\!\!\!\!-\sqrt{(r{-}1)(r{+}k{-}1)} 
                       & 2r+k-1{\mbf{\,-\,r}} 
\end{pmatrix}
\end{align}
\begin{align} \label{QkerP}
\sfrac{1}{2} H^{(k)}_{\text{ker}P} \= \begin{pmatrix}
2r+k+1{\mbf{\,-\,r}} & -\sqrt{(r{+}1)(r{+}k{+}1)} & & & \\[8pt]
-\sqrt{(r{+}1)(r{+}k{+}1)} & 2r+k+3 & -\sqrt{(r{+}2)(r{+}k{+}2)} & & \\[8pt]
 & -\sqrt{(r{+}2)(r{+}k{+}2)} & 2r+k+5 & \qquad\ddots\!\!\!\!\!\!\!\! & \\[8pt]
 & & \!\!\!\!\!\!\!\!\!\!\!\!\!\!\!\!\ddots & \ddots & & \\[8pt]
\end{pmatrix} \ ,
\end{align}
where the boldface contributions disturb the systematics and originate
from the $\Phi\Delta\Phi^\+$ term in the Hessian.
The previous paragraph asserts strict positivity of~$H$ for $k>r$.
Due to the finiteness of $H^{(k)}_{\text{Gr}(P)}$, the Hessian 
on $\Ecal_{k>r}$ features precisely $r$ positive eigenvalues\footnote{
meaning that the corresponding modes are normalizable eigenvectors,
i.e.~Hilbert-Schmidt}    
in its spectrum. 
We shall argue below that $H^{(k)}_{\text{ker}P}$ contributes a purely 
continuous spectrum~$\R_+$ for any~$k$. Thus, the above eigenvalues are
not isolated but imbedded in the continuum.

It is instructive to look at the edge of the continuum. The non-normalizable 
zero mode of $H^{(k)}_{\text{Gr}(P)}$ is explicitly given by
\begin{equation} \label{edge}
\mu_{r+m}^{\text{zero}} \= \mu_r\sqrt{\frac{
(r{+}k{+}1)(r{+}k{+}2)\cdots(r{+}k{+}m)}{(r{+}1)(r{+}2)\cdots(r{+}m)}}
\= \mu_r\sqrt{\frac{
(r{+}m{+}1)(r{+}m{+}2)\cdots(r{+}m{+}k)}{(r{+}1)(r{+}2)\cdots(r{+}k)}} \ ,
\end{equation}
which grows like $m^{k/2}$ when $m=0,1,2,\dots$ gets large.
Being unbounded for $k{>}0$ this infinite vector does not yield an admissible
perturbation of~$\Phi_r$ however: $\delta^2 E$ is infinite. 

\subsubsection{Slightly off-diagonal perturbations}
\noindent
For each value of $k$ in the range $1\le k\le r$ we already established 
in Section~3.3 the existence of an odd complex normalizable zero mode
connected with a moduli parameter. Nevertheless, as we will show now,
$E^{(2)}[\Phi_r,\phi_{(k)}]$ remains positive for $1\le k\le r$ albeit 
not strictly so. In order to simplify (\ref{QPhik}) we make use of
the following property for $R_j(\mu)$ as defined in~(\ref{Rmdef}):
\begin{equation} \label{Rid}
\begin{aligned}
\sum_{j=m}^{l-1}R_j(\mu)
&\= k\,|\mu_l|^2\ -\ k\,|\mu_m|^2\ +\ 2\sum_{j=m}^{l-1}
|\sqrt{j{+}1}\,\mu_{j+1}-\sqrt{j{+}k{+}1}\,\mu_{j}|^2 \\
&\= k\,|\mu_m|^2\ -\ k\,|\mu_l|^2\ +\ 2\sum_{j=m}^{l-1}
|\sqrt{j{+}k{+}1}\,\mu_{j+1}-\sqrt{j{+}1}\,\mu_{j}|^2\ .
\end{aligned}
\end{equation}
Here, by definition, all integers are non-negative and all sums are taken
to vanish if $m\ge l$. The two equations above are easily proved by induction
over $l$, starting from the trivial case $l=m$.
We now employ the algebraic identities~(\ref{Rid}) to rewrite~(\ref{QPhik})
for $1\le k\le r$ and obtain
\begin{align} \label{QPhik3}
\sfrac{1}{2\pi} E^{(2)}[\Phi_r,\phi_{(k)}] &\=
k\,|\mu_r|^2\ +\ \sum_{j=r}^{\infty}R_j(\mu) \\
&\quad +\ 2\sum_{j=0}^{r-k-2}
|\sqrt{j{+}1}\,\mu_{j+1}-\sqrt{j{+}k{+}1}\,\mu_{j}|^2
\ +\ 2\!\sum_{j=r-k}^{r-2}
|\sqrt{j{+}k{+}1}\,\mu_{j+1}-\sqrt{j{+}1}\,\mu_{j}|^2\ , \nonumber 
\end{align}
which is indeed positive semi-definite. 
We observe that the even part of~$E^{(2)}$ now consists of two disjoint pieces,
containing $\{\mu_0,\dots,\mu_{r-k-1}\}$ and $\{\mu_{j\ge r}\}$,
which are separated by the odd perturbations $\{\mu_{r-k},\dots,\mu_{r-1}\}$.
If the right-hand side of~(\ref{QPhik3}) is equal 
to zero, then all components $\mu_{j\ge r}$ must vanish because each $R_j$
is a strictly positive quadratic form of $\mu_j$ and $\mu_{j+1}$. In contrast,
the remaining coefficients $\mu_{j<r}$ need not vanish at the zero set 
of $E^{(2)}[\Phi_r,\phi_{(k)}]$. In order to find the zero modes of
the Hessian, it is convenient to visualize it again in matrix form, namely
\begin{equation}
H^{(k)} \= H^{(k)}_{\text{im}P}\,
\oplus\,H^{(k)}_{\text{Gr}(P)}\,\oplus\,H^{(k)}_{\text{ker}P} \ ,
\end{equation}
where the blocks have sizes $r{-}k$, $k$, and $\infty$, respectively.
{}From (\ref{QPhik3}) we extract
\goodbreak
\begin{align}
\sfrac{1}{2} H^{(k)}_{\text{im}P} \= \begin{pmatrix}
k+1 & -\sqrt{1(k{+}1)} & & & & \\[8pt]
-\sqrt{1(k{+}1)} & k+3 & -\sqrt{2(k{+}2)} & & & \\[8pt]
 & -\sqrt{2(k{+}2)} & k+5 & \qquad\ddots\!\!\!\!\!\!\!\! & & \\[8pt]
 & & \!\!\!\!\!\!\!\!\ddots & \;\ddots & \ddots & \\[8pt]
 & & & \!\!\!\!\!\!\!\!\!\!\!\!\!\!\!\!\ddots 
                    & 2r-k-3 & -\sqrt{(r{-}k{-}1)(r{-}1)} \\[8pt]
 & & & & \!\!\!\!\!\!\!\!\!\!\!\!-\sqrt{(r{-}k{-}1)(r{-}1)} 
                    & 2r-k-1{\mbf{\,-\,r}} 
\end{pmatrix}
\end{align}
%\vskip-5mm
\begin{align}
\sfrac{1}{2} H^{(k)}_{\text{Gr}(P)} \= \begin{pmatrix}
2r-k+1{\mbf{\,-\,r}} & -\sqrt{(r{-}k{+}1)(r{+}1)} & & & \\[8pt]
-\sqrt{(r{-}k{+}1)(r{+}1)} & 2r-k+3 & \qquad\ddots\!\!\!\!\!\!\!\! & & \\[8pt]
 & \!\!\!\!\!\!\!\!\!\!\!\!\!\!\!\!\!\!\!\!\ddots & \!\!\!\!\!\!\!\!\;\ddots 
                   & \qquad \ddots & & \\[8pt]
 & & \!\!\!\!\!\!\!\!\!\!\!\!\!\!\!\!\!\!\!\!\!\!\!\!\!\!\!\!\!\!\!\!\!\!\!\!
     \!\!\!\!\!\!\!\!\ddots & \!\!\!\!\!\!\!\!\!\!\!\! 2r+k-3 
                   & \!\!\!\!-\sqrt{(r{-}1)(r{+}k{-}1)} \\[8pt]
 & & 
 & \!\!\!\!\!\!\!\!\!\!\!\!\!\!\!\!\!\!\!\!\!\!\!\!-\sqrt{(r{-}1)(r{+}k{-}1)}
                   & \!\!\!\ 2r+k-1{\mbf{\,-\,r}} 
\end{pmatrix}
\end{align}
and $H^{(k)}_{\text{ker}P}$ being identical to the matrix in~(\ref{QkerP}).
The extreme cases are $\ H^{(1)}_{\text{Gr}(P)}=(0)\ $ and 
$\ H^{(r-1)}_{\text{im}P}=(0)$, while $H^{(r)}_{\text{im}P}$ is empty.

We already know that $H^{(k)}_{\text{ker}P}$ is strictly positive definite. 
As claimed earlier, its spectrum is $\R_+$ and purely continuous,
with a non-normalizable zero mode given by~(\ref{edge}). 
Being finite-dimensional, the other two blocks jointly yield again just $r$ 
non-negative and non-isolated proper eigenvalues for fixed~$k{>}0$. 
Interestingly, two of these eigenvalues are now zero and give us one even and 
one odd complex normalizable zero mode. With $\beta_k,\gamma_k\in\C$ the latter
take the following form:\footnote{
The even zero mode is of course absent for $k=r$.}
\goodbreak
\begin{align} \label{evenzero}
H^{(k)}_{\text{im}P}:\qquad
(\mu_\ell)_{\text{zero}} &\= \gamma_k\,\begin{pmatrix}
\qquad\ \;\sqrt{1}\sqrt{2}\sqrt{3} \cdots 
                    \sqrt{r{-}k{-}2}\sqrt{r{-}k{-}1} \\[4pt]
\quad\ \sqrt{k{+}1}\sqrt{2}\sqrt{3} \cdots 
                    \sqrt{r{-}k{-}2}\sqrt{r{-}k{-}1} \\[4pt]
\sqrt{k{+}1}\sqrt{k{+}2}\sqrt{3} \cdots 
                    \sqrt{r{-}k{-}2}\sqrt{r{-}k{-}1} \\[-2pt]
\vdots \\[2pt]
\sqrt{k{+}1}\sqrt{k{+}2}\sqrt{k{+}3} \cdots 
                    \sqrt{r{-}2}\sqrt{r{-}k{-}1} \\[4pt]
\sqrt{k{+}1}\sqrt{k{+}2}\sqrt{k{+}3} \cdots 
                    \sqrt{r{-}2}\sqrt{r{-}1} \quad\ 
\end{pmatrix} \ , \\[10pt] \label{oddzero}
H^{(k)}_{\text{Gr}(P)}:\qquad
(\mu_\ell)_{\text{zero}} &\= \beta_k\,\begin{pmatrix}
\qquad\; \sqrt{r{+}1}\sqrt{r{+}2}\sqrt{r{+}3} \cdots 
                    \sqrt{r{+}k{-}2}\sqrt{r{+}k{-}1}\\[4pt]
\quad\ \sqrt{r{-}k{+}1}\sqrt{r{+}2}\sqrt{r{+}3} \cdots 
                    \sqrt{r{+}k{-}2}\sqrt{r{+}k{-}1}\\[4pt]
\sqrt{r{-}k{+}1}\sqrt{r{-}k{+}2}\sqrt{r{+}3} \cdots
                    \sqrt{r{+}k{-}2}\sqrt{r{+}k{-}1}\\[-2pt]
\vdots \\[2pt]
\sqrt{r{-}k{+}1}\sqrt{r{-}k{+}2}\sqrt{r{-}k{+}3} \cdots
                    \sqrt{r{-}2}\sqrt{r{+}k{-}1}\\[4pt]
\sqrt{r{-}k{+}1}\sqrt{r{-}k{+}2}\sqrt{r{-}k{+}3} \cdots
                    \sqrt{r{-}2}\sqrt{r{-}1} \quad\
\end{pmatrix} \ .
\end{align}
Altogether, there are $2r{-}2$ real zero modes
in $H_{\text{im}P}$ and $2r$ real zero modes in $H_{\text{Gr}(P)}$. 
Identifying $\mu_{(k)j}=\Lambda_{j,k+j}=-\bar\Lambda_{k+j,j}$,
the latter precisely agree with the BPS moduli found in~(\ref{Lsol}).
The former zero modes correspond to moduli outside the Grassmannian, thus 
generating nearby non-diagonal non-BPS solutions to the equation of motion.

\subsubsection{Diagonal perturbations}
\noindent
We come to the diagonal (or radial) perturbations $\phi_{(0)}\in\Ecal_0$
(see~(\ref{phi0})). The transformation of 
\begin{equation}
\sfrac{1}{2\pi} E^{(2)}[\Phi_r,\phi_{(0)}] \= 
\sum_{m=0}^\infty (m{+}1)\,(\phi_{m+1}-\phi_m)^2
-2r\,( \phi_r^2 + \phi_{r-1}^2 ) 
\end{equation}
to a sum of squares is much easier than that of~(\ref{QPhik}),
but the result shows one minus in the signature:
\begin{equation} \label{QPhi0}
\sfrac{1}{2\pi} E^{(2)}[\Phi_r,\phi_{(0)}] \= -r\,(\phi_{r-1}+\phi_{r})^2\ 
+\!\sum_{j\ge0,\,j\neq r-1}\!\!(j{+}1)\,(\phi_{j+1}-\phi_j)^2\ .
\end{equation} 
However, we can use this expression to conclude that the second variation
form $E^{(2)}$ cannot have more than one negative mode on $\Ecal_0$.

{\bf Lemma}.
{\it Let $V\subset\Ecal_0$ be a real vector subspace such that
$E^{(2)}[\Phi,\phi]<0$ for all nonzero $\phi\in V$. Then $V$ is at
most one-dimensional. In other words, one cannot find linearly
independent vectors $\phi,\psi\in\Ecal_0$ such that $E^{(2)}[\Phi,.]$ takes
negative values on each non-zero linear combination of $\phi$ and~$\psi$.

Proof}. A necessary and sufficient condition for a pair of linearly
independent vectors $\phi,\psi\in\Ecal_0$ to span a negative subspace
for $E^{(2)}(.):=E^{(2)}[\Phi,.]$ is that
\begin{equation}
E^{(2)}(\phi)<0\ ,\quad E^{(2)}(\psi)<0 \quad\text{and}\quad 
B(\phi,\psi)^2-E^{(2)}(\phi)E^{(2)}(\psi)<0 \ ,
\end{equation}
where 
$B(\phi,\psi):=\frac12\{E^{(2)}(\phi{+}\psi)-E^{(2)}(\phi)-E^{(2)}(\psi)\}$ is
the corresponding bilinear form. If $\phi,\psi\in\Ecal_0$
satisfy these inequalities, then the same holds for their truncations
(that is, vectors obtained by replacing all the coordinates
$\phi_j$, $\psi_j$ with $j\ge n$ by zero) if $n$ is large enough.
Thus, we may assume the existence of $n$ such
that $\phi_j=\psi_j=0$ for $j\ge n$. But the restriction of $E^{(2)}$
to the space $\R^n$ of all vectors
$\phi=(\phi_0,\phi_1,\dots,\phi_{n-1},0,0,\dots)$
has signature $(1,n{-}1)$ according to~(\ref{QPhi0}) and, therefore,
this space contains no two-dimensional negative subspaces.\qed

For an alternative visualization, we write the second-order fluctuation form as
\begin{equation}
E^{(2)}[\Phi_r,\phi_{(0)}]\=
2\pi\sum_{m,\ell=0}^\infty \phi_m\,H^{(0)}_{m\ell}\,\phi_\ell\ ,
\end{equation}
where the matrix entries $H^{(0)}_{m\ell}$ can be extracted from (\ref{Hdiag}) 
or from (\ref{QPhi0}) as
\begin{equation}
\sfrac{1}{2\pi}\,H^{(0)}_{m\ell}
\= (2m{+}1-2r\delta_{m,r-1}-2r\delta_{m,r})\,\delta_{m,\ell}
- m\,\delta_{m,\ell+1} - (m{+}1)\,\delta_{m+1,\ell} \ .
\end{equation}
In matrix form this reads as ($m,\ell\in\NN_0$)
\begin{equation} \label{Qzero}
\begin{aligned}
(H^{(0)}_{m\ell}) &\= \begin{pmatrix} 
\phantom{-}1 &           -1 \\[2pt]
          -1 & \phantom{-}3 &           -2 \\[2pt]
             &           -2 & \phantom{-}5 & -3 \\[-2pt]
             &              &           -3 & \phantom{-}7 & \ddots \\[-2pt]
             &              &              & \ddots       & \ddots
\end{pmatrix}
\ -\ 2\,r\,(\delta_{m,r-1}\delta_{\ell,r-1}\,+\,\delta_{m,r}\delta_{\ell,r})
\\[8pt]
&\= \begin{pmatrix} 
\ddots & \ddots & & & & \\[-2pt]
\ddots & 2r{-}3 & -r{+}1 & & & \\[2pt]
& -r{+}1 & \mbf{-1} & -r & & \\[2pt]
& & -r & \mbf{+1} & -r{-}1 & \\[-2pt]
& & & -r{-}1 & 2r{+}3 & \ddots \\[-2pt]
& & & & \ddots & \ddots 
\end{pmatrix} \ .
\end{aligned}
\end{equation}
In distinction to the earlier cases, this infinite Jacobian (i.e.~tri-diagonal)
matrix does not split. We shall compute its spectrum in a moment.
Prior to this, some observations can already be made.

First of all, the unique zero mode of $H^{(0)}$ is found from (\ref{edge})
for $k{=}0$ and given by ($\gamma_0\in\R$)
\begin{equation}
\phi_{\text{zero}}\=\ic\gamma_0\,\Phi_r
\qquad\Longleftrightarrow\qquad  (\phi_m)_{\text{zero}} \= \gamma_0\,
(\underbrace{-1,\ldots,-1}_{r\:\text{times}},+1,+1,\ldots) \ ,
\end{equation}
which clearly generates the global phase rotation symmetry
$\Phi\to\e^{\ic\gamma_0}\Phi$ already noted in~(\ref{globalsym}).
In contrast to the zero modes in~$\Ecal_{k>0}$ depicted in (\ref{edge}),
(\ref{evenzero}) and~(\ref{oddzero}), $\phi_{\text{zero}}$ is not 
Hilbert-Schmidt but bounded. Although it does not belong to a proper zero 
eigenvalue of the Hessian, $\phi_{\text{zero}}$, being proportional 
to~$\Phi_r$, still meets all our requirements for an admissible perturbation.

In order to identify unstable modes we turn on all diagonal perturbations
in a particular manner suggested by~(\ref{path}),
\begin{equation}
\Phi(\{\alpha_m\}) \= \unity\ -\
\sum_{m=0}^{r-1} (\e^{\ic\alpha_m}{+}1)\,|m\>\<m| \ +\
\sum_{m=r}^\infty  (\e^{\ic\alpha_m}{-}1)\,|m\>\<m| \ .
\end{equation}
The energy of this configuration is easily calculated to be
\begin{equation}
\sfrac{1}{8\pi}\,E(\{\alpha_m\}) \=
\sin^2\sfrac{\alpha_0-\alpha_1}{2} +\dots
+(r{-}1)\sin^2\sfrac{\alpha_{r-2}-\alpha_{r-1}}{2}
+r\cos^2\sfrac{\alpha_{r-1}-\alpha_r}{2}
+(r{+}1)\sin^2\sfrac{\alpha_r-\alpha_{r+1}}{2} +\dots
\end{equation}
with a single $\cos^2$ appearing only in the indicated place.
Expanding up to second order in the real parameters~$\alpha_m$ and
putting $\alpha_m=0$ for $m\ge r$, we find that
\begin{equation}
H^{(0)}<0 \qquad\Longleftrightarrow\qquad
(\alpha_0{-}\alpha_1)^2 + 2(\alpha_1{-}\alpha_2)^2 + \dots +
(r{-}1)(\alpha_{r-2}{-}\alpha_{r-1})^2 - r\,\alpha_{r-1}^2 \ \le\ 0 \ ,
\end{equation}
which determines a convex cone in the $r$-dimensional restricted fluctuation
space. The most strongly negative mode is given by ($\alpha\in\R$)
\begin{equation}
\phi_{\text{neg}}\=-\ic\alpha P_r
\qquad\Longleftrightarrow\qquad (\phi_m)_{\text{neg}} \= -\alpha\,
(\underbrace{1,\ldots,1}_{r\:\text{times}},0,0,\ldots) \ .
\end{equation}
Comparison with (\ref{Epath}) reveals that following this mode one arrives at
$\Phi=\unity$. More generally,
\begin{equation}
(\phi_m) \= -\alpha\,
(\underbrace{0,\ldots,0}_{\tilde r\:\text{times}},
 \underbrace{1,\ldots,1}_{(r-\tilde r)\:\text{times}},0,0,\ldots)
\end{equation}
perturbs in the direction of $\Phi=\unity-2P_{\tilde r}$.
None of these modes are eigenvectors of the matrix~$H^{(0)}$.

It is instructive to confirm these assertions numerically.
To this end, we truncate the matrix $(H^{(0)}_{m\ell})$ at some cut-off value
$m_{\text{max}}=\ell_{\text{max}}$ and diagonalize the resulting
finite-dimensional matrix. Truncating at values $m_{\text{max}}<r$ only
allows for positive eigenvalues, whereas for values $m_{\text{max}}\geq r$
we obtain exactly {\em one\/} negative eigenvalue.
Its numerical value depends roughly linearly on $r$ and converges very quickly
with increasing cut-off as can be seen from the following graphs.
\vskip-5mm
\begin{figure}[H]
\psfrag{$k_max$}{$\scriptstyle{m_{\text{max}}}$}
\psfrag{$lambda$}{$\scriptstyle{\lambda}$}
\begin{center}
\includegraphics[origin=ct,width=80mm]{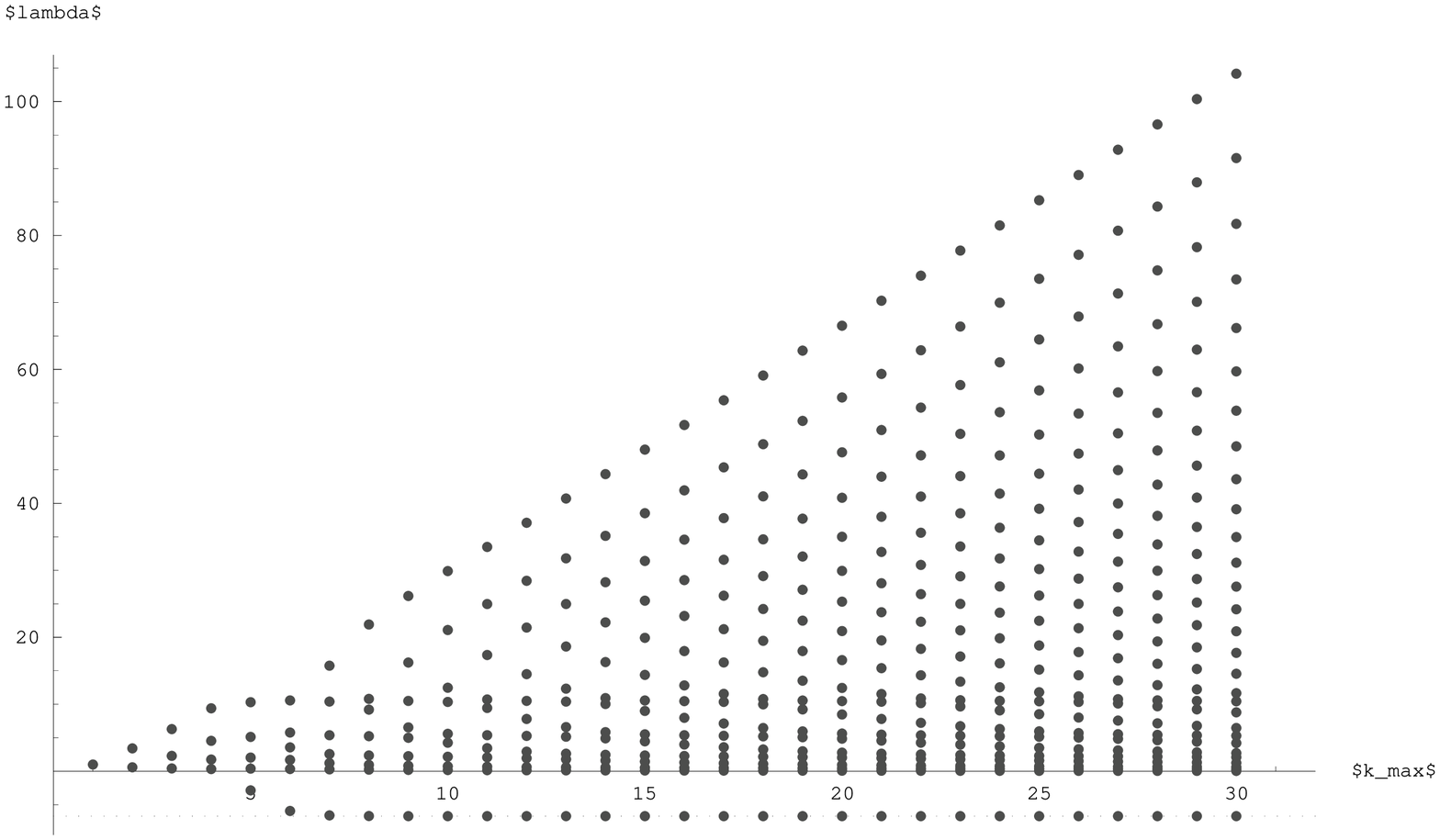}
\hspace{\fill}
\includegraphics[origin=ct,width=80mm]{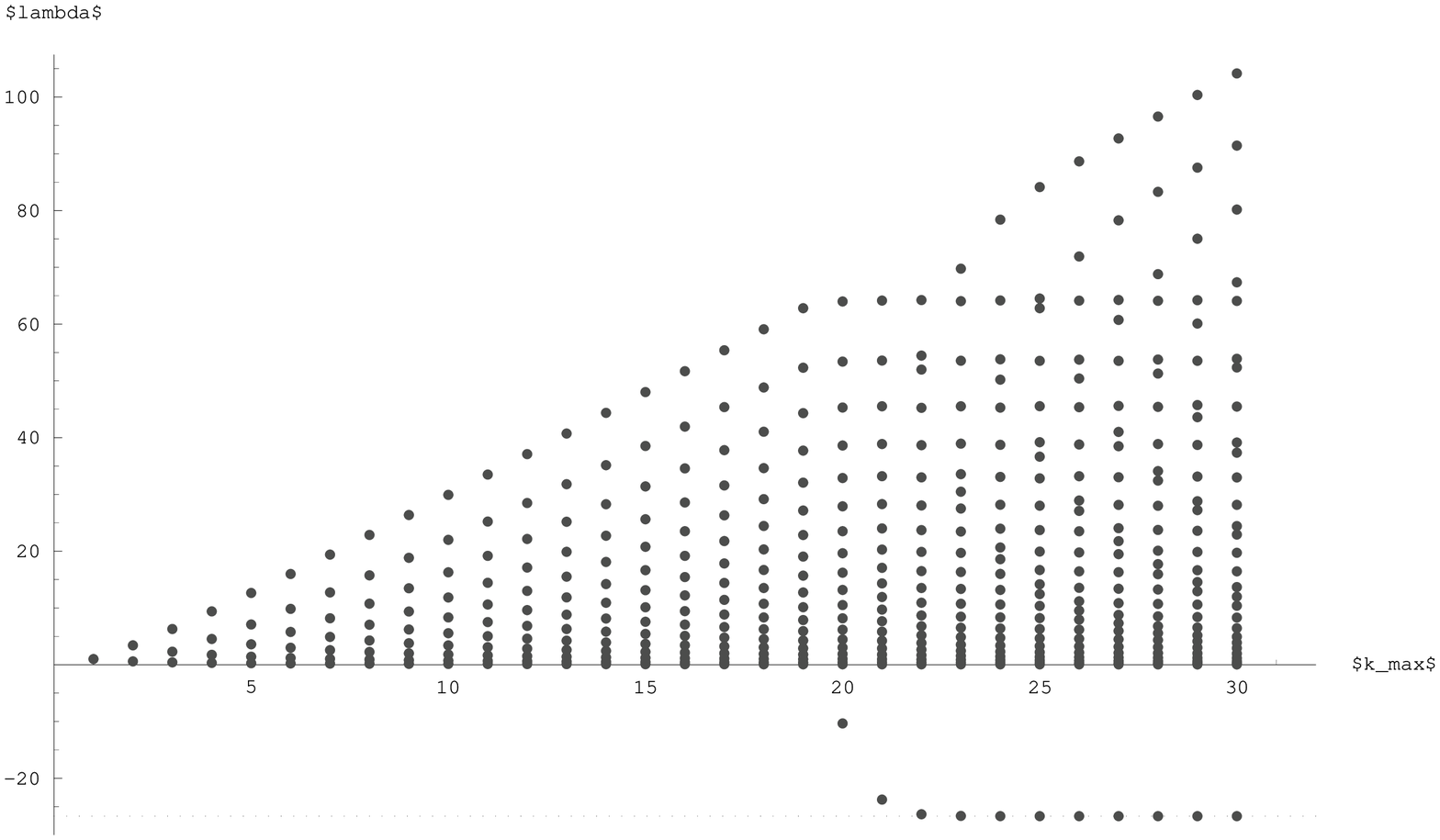}
\end{center}
\caption{Possible eigenvalues $\lambda$ versus the cut-off parameter
$m_{\text{max}}$ for $r=5$ and $r=20$}
\end{figure}

\subsubsection{Spectrum of the Hessian}
\noindent
The spectrum of the restriction $H^{(k)}=\Delta_k-(\Phi_r\Delta\Phi_r)_k$
of the Hessian $H$ to $\Ecal_k$ can actually be calculated by
passing to a basis of Laguerre polynomials. The following considerations
are based on spectral theory and require all vectors to be in $l_2$,
i.e.~all operators to be Hilbert-Schmidt.

Let us first consider~$\Ecal_0$.
We use the following property of the noncommutative
Laplacian $-\Delta_0$: sending each basis vector $|m\>$ of $\Hcal$ to the
$m$-th Laguerre polynomial $L_m(x)$, we get a unitary map
$U:\Hcal\to L^2(\R_+,\e^{-x}\diff x)$ such that $U\Delta_0U^{-1}$ is just
the operator of multiplication by~$x$: 
\begin{equation} \label{xL}
x\,L_m(x) \= -m\,L_{m-1}(x) + (2m{+}1)L_m(x) - (m{+}1)L_{m+1}(x) \ ,
\end{equation}
which is to be compared with (\ref{Qzero}).
Hence, in the basis of the Laguerre polynomials, the eigenvalue equation
$H^{(0)}|\phi\>=\lambda|\phi\>$ is rewritten in terms of $f:=U|\phi\>$ as
\begin{equation} \label{ev}
x\,f(x)-2r\bigl(\<f,L_{r-1}\>L_{r-1}(x)+\<f,L_r\>L_r(x)\bigr)\=\lambda\,f(x)\ .
\end{equation}
Clearly, a function $f\in L^2(\R_+,\e^{-x}\diff x)$ satisfies (\ref{ev})
if and only if it is given by
\begin{equation} \label{fsol}
f(x)\=\frac{c_0L_{r-1}(x)+c_1L_r(x)}{x-\lambda}
\end{equation}
and the constant coefficients $c_0,c_1\in\R$ satisfy the linear system
\begin{equation} \label{linsys}
\begin{cases} 
\left(I_{r-1,r-1}(\lambda)-\frac1{2r}\right)c_0+I_{r-1,r}(\lambda)\,c_1\=0
\\[6pt]
I_{r,r-1}(\lambda)\,c_0+\left(I_{r,r}(\lambda)-\frac{1}{2r}\right)c_1\=0
\end{cases} \ ,
\end{equation}
which is obtained simply by inserting (\ref{fsol}) into (\ref{ev}).
Here we use the notation
\begin{equation} \label{ints}
I_{k,l}(\lambda)\ :=\ 
\int_0^{\infty} \frac{\e^{-x}\,\diff x}{x-\lambda}\; L_k(x)\,L_l(x)
\qquad\text{for}\quad k,l\ge0\ . 
\end{equation}

The integrals (\ref{ints}) are simple (though not elementary)
special functions of $\lambda$. Indeed,
$I_{00}(\lambda)$ is a version of the integral logarithm:
$I_{00}(\lambda)=-\e^{-\lambda}\operatorname{li}(\e^{\lambda})$ 
for all $\lambda<0$. On the other hand, using the recursion relations
\begin{equation}    
(k{+}1)\,L_{k+1}(x)\=(2k{+}1{-}x)\,L_k(x)\ -\ k\,L_{k-1}(x)\ , 
\end{equation}
one can show by induction over $k$ and $l$ that all functions
$I_{k,l}(\lambda)$ are expressed in terms of $I_{00}(\lambda)$: 
\begin{equation}
I_{kl}(\lambda)\=A_{kl}(\lambda)\,I_{00}(\lambda)\ +\ B_{kl}(\lambda)\ , 
\end{equation}
where $A_{kl}$ and $B_{kl}$ are polynomials in $\lambda$ of degree at 
most~$k{+}l$. Hence, the determinant
\begin{equation}
F_r(\lambda)\ :=\ \left\|\begin{matrix}
I_{r-1,r-1}(\lambda)-\frac{1}{2r}  & I_{r-1,r}(\lambda) \\[6pt]
I_{r,r-1}(\lambda) & I_{r,r}(\lambda)-\frac{1}{2r} \end{matrix}\right\|
\end{equation}
of the linear system (\ref{linsys}) is a known special function
of $\lambda$, whose zeros $\lambda_r$ on the negative semiaxis are precisely
the negative eigenvalues of the Hessian operator $H^{(0)}[\Phi_r]$ for the
diagonal BPS background of rank~$r$.

One can now prove the existence of negative eigenvalues. 
By verifying that the real numbers 
$F_r(-\infty):=\lim_{\lambda\to-\infty}F_r(\lambda)$ and
$F_r(0):=\lim_{\lambda\to0-} F_r(\lambda)$ have different signs,
we can conclude that $F_r(\lambda)$ possesses at least one zero on the
negative semiaxis.
For example, take $r=1$. In this case we find
\begin{equation}
\left.\begin{matrix}
I_{01}(\lambda)\= I_{10}(\lambda) & \!=\ (1{-}\lambda)I_{00}(\lambda)-1 
\hfill \\[6pt] \hfill
I_{11}(\lambda) & \!=\ (1{-}\lambda)^2 I_{00}(\lambda)+\lambda-1
\end{matrix}
\right\}\quad\Longrightarrow\qquad
2\,F_1(\lambda)\=-\lambda-\sfrac12-\lambda^2 I_{00}(\lambda) \ .
\end{equation}
Passing to the limit under the integral sign, we see that
$F_1(-\infty)=\frac14$ and $F_1(0)=-\frac14$. This proves the existence 
of a negative eigenvalue of $H^{(0)}[\Phi_1]$.

Can $H^{(0)}[\Phi_r]$ also have non-negative eigenvalues? To disprove this 
possibility, we consider the eigenvalue equation~(\ref{ev}) and show that 
it admits only the trivial solution $f(x)\equiv 0$ if $\lambda\ge 0$.
Indeed, for non-negative~$\lambda$ the solution~(\ref{fsol}) is not 
square-integrable unless
\begin{equation}
c_0L_{r-1}(\lambda)+c_1L_{r}(\lambda)\=0 \qquad\Longrightarrow\qquad
c_0\=c\,L_{r}(\lambda) \qquad\text{and}\qquad  c_1\=-c\,L_{r-1}(\lambda)
\end{equation}
for some constant~$c$. 
Therefore, the solution~(\ref{fsol}) is completely determined as
\begin{equation} \label{fsol2}
f(x)\=c\,\frac{L_r(\lambda)L_{r-1}(x)-L_{r-1}(\lambda)L_r(x)}{x-\lambda}
\=c\,\sum_{k=0}^{r-1} L_k(\lambda)\,L_k(x)
\end{equation}
via the Christoffel-Darboux formula. 
It follows that $f$ is orthogonal to $L_{r}$. Inserting $\<f,L_r\>=0$
into~(\ref{ev}), we learn that the polynomial $(x{-}\lambda)f(x)$ is 
proportional to $L_{r-1}(x)$. But then the first equality in~(\ref{fsol2}) 
demands that $L_{r-1}(\lambda)=0$, constraining~$\lambda$. Feeding this
into the second expression for~$f$ in~(\ref{fsol2}) we see that the sum
runs to $r{-}2$ only, implying that $f$ is orthogonal to $L_{r-1}$ as well.
This finally simplifies the eigenvalue equation~(\ref{ev}) to the one for
$\Delta_0$, namely
\begin{equation}
x\,f(x) \= \lambda\,f(x) \ ,\qquad\text{whence}\qquad
f(x)\ \equiv\ 0 \ ,
\end{equation}
as required. 
Hence, the non-negative part of the spectrum of~$H^{(0)}$ is purely continuous.

A similar analysis can be applied to the Hessian on~$\Ecal_{k>0}$.
In this case, the appropriate polynomials are the normalized (generalized)
Laguerre polynomials~$L^k_m$, which form an orthonormal basis for
$L^2(\R_+,x^k\e^{-x}\diff{x})$. One rediscovers the $r$ proper eigenvalues
as zeroes of the characteristic polynomial for 
$H^{(k)}_{\text{im}P}\oplus H^{(k)}_{\text{Gr}(P)}$, 
accompanied by a continuous spectrum covering the positive semiaxis for
$H^{(k)}_{\text{ker}P}$, as claimed earlier.

If we do not care for resolving non-isolated eigenvalues, we may also infer
the spectrum of the Hessian~$H$ from the classical Weyl theorem
(see~\cite{Kato}, Theorem IV.5.35 and \cite{RS}, \S~XIII.4, Example~3).

{\bf Assertion}.
{\it Let $\Phi=\Phi_r$ be the diagonal\/} BPS {\it solution of rank~$r>0$. Then
\begin{equation} \label{assert}
\operatorname{spectrum}\,(H[\Phi_r])\=\{\lambda_r\}\cup\,[0,+\infty)\ ,
\end{equation}
where $\lambda_r$ is a negative eigenvalue of multiplicity~$1$ and 
$[0,+\infty)$ is the essential spectrum.\footnote{
We recall that the essential spectrum of a self-adjoint operator is the whole 
spectrum minus all isolated eigenvalues of finite multiplicity. 
The multiplicity of an eigenvalue is the dimension of the corresponding 
eigenspace.} 
Furthermore, we have $-2r<\lambda_r<0$.

Proof}. Weyl's theorem asserts that the essential spectrum is preserved
under compact perturbations. The essential spectrum of $\Delta$ (on all
subspaces~$\Ecal_k$) is known to be equal to $[0,+\infty)$. (This follows, 
for example, from the explicit diagonalization of $\Delta_k$ in terms of 
the Laguerre polynomials, as indicated above.)
Since the perturbation $(\Phi_r\Delta\Phi_r)$ is finite-dimensional, 
Weyl's theorem applies to show that the essential spectrum of~$H$ is also 
equal to $[0,+\infty)$. Although we have explicitly shown that $H^{(k)}$ is
positive semi-definite for~$k{>}0$, the essential spectrum~$\R_+$ cannot 
exhaust the whole spectrum because the operator~$H^{(0)}$ is not positive
semi-definite on $\Ecal_0$. (Indeed, (\ref{QPhi0}) shows that the quadratic 
form~$E^{(2)}$ is negative on many finite vectors.) Hence, $H^{(0)}$ must 
have negative eigenvalues of finite multiplicity.
The above lemma implies that there can be only one such eigenvalue, and
its multiplicity must be equal to~1. This proves (\ref{assert}) and the
inequality $\lambda_r<0$. The estimate $-2r<\lambda_r$ follows because the
norm of the operator $(\Phi_r\Delta\Phi_r)$ is equal to $2r$, so 
$H[\Phi_r]+2r\unity_\Hcal\ge\Delta$ is non-negative definite and hence
cannot have negative eigenvalues.\qed

\subsection{Results for non-diagonal U(1) BPS backgrounds}
\noindent
Even though any non-diagonal BPS background~$\Phi$ can be reached from a
diagonal one by a unitary transformation~$U$ which does not change
the value of the energy, the fluctuation problem will get modified 
under such a transformation because
\begin{equation}
U\,a\,U^\+ \= f(a,\adag) \qquad\text{and}\qquad
U\,\adag U^\+ \= f^\+(a,\adag)
\end{equation}
will in general not have an action as simple as $a$ and $\adag$ in the
original oscillator basis. 
The exceptions are the symmetry transformations, of which
only the rigid translations $D(\a)$ and rotations $R(\vartheta)$ 
keep $\Phi$ within the Grassmannian.
To be specific, we recall that
\begin{equation}
\begin{aligned}
D(\a)^\+\,a\,D(\a) \= a + \a 
&\qquad\Longrightarrow\qquad&
E\,[D(\a)\,\Phi\,D(\a)^\+] \= E\,[\Phi] \ ,\\[4pt]
R(\vartheta)^\+\,a\,R(\vartheta) \= \e^{\ic\vartheta} a 
&\qquad\Longrightarrow\qquad&
E\,[R(\vartheta)\,\Phi\,R(\vartheta)^\+] \= E\,[\Phi] \ .
\end{aligned}
\end{equation}
The invariance of the spectrum of~$H$ under translations (or rotations)
can also be seen directly:
\begin{equation}
\begin{aligned}
H\,\phi_n &\= \epsilon_n\,\phi_n \qquad\text{and}\qquad
\Delta\,(D\,f\,D^\+) \= D\,(\Delta f)\,D^\+ \quad\Longrightarrow \\[4pt]
H'\phi_n' &\ \equiv\
\bigl[ \Delta-(D\Phi D^\+)\Delta(D\Phi D^\+) \bigr] D\phi_n D^\+ \=
D\,\bigl([\Delta-\Phi\Delta\Phi]\phi_n \bigr)D^\+ \=\epsilon_n\,\phi_n' \ .
\end{aligned}
\end{equation}
More general unitary transformations will not simply commute with the action
of $\Delta$ and thus change the spectrum of~$H$.

In the rank-one BPS situation, any solution is a translation of~$\Phi_1$,
and hence our previous discussion of fluctuations around diagonal U(1) BPS
backgrounds completely covers that case. This is no longer true for higher-rank
BPS backgrounds. For a complete stability analysis of abelian $r$-solitons 
it is therefore necessary to investigate separately the spectrum of~$H$ 
around each soliton configuration (\ref{Pcoherentr}) based on $r$~coherent 
states, whose center of mass may be chosen to be the origin. Since the
background holomorphically depends on $r$ parameters~$\a_1,\dots,\a_r$ and
passes to $\Phi_r$ when all parameters go to zero, one may hope to show
that the spectrum of~$H$ does not change qualitatively when $\Phi$ varies
inside the rank-$r$ moduli space.

\section{Perturbations of U(2) backgrounds}
\noindent 
In this section we investigate the behavior of a simple U(2) BPS
solution under perturbations and conclude that the fluctuation analysis
reduces to the U(1) case discussed in the previous section.

\subsection{Results for diagonal U(2) BPS backgrounds}
\noindent
We consider a nonabelian BPS projector of the diagonal type considered 
in \eqref{simplespecialP}:
\begin{equation}
P \= \unity_\Hcal\,\oplus\,P_Q \qquad\text{where}\qquad
P_Q \= \sum_{k=0}^{Q-1} |k\>\<k|\ .
\end{equation}
Clearly this describes a BPS solution 
\begin{equation}
\Phi \= \begin{pmatrix} \Phi^{(1,1)} & \Phi^{(1,2)} \\[4pt]
                        \Phi^{(2,1)} & \Phi^{(2,2)} \end{pmatrix}
     \= \begin{pmatrix} - \unity_\Hcal & 0\\[4pt]
                               0       & \unity_\Hcal-2P_Q \end{pmatrix}
\end{equation}
of energy $E=8 \pi Q$. 
Inserting this into the expression (\ref{Qform}) for the quadratic
energy correction~$E^{(2)}$ one obtains
%\goodbreak
\begin{equation}
\begin{aligned}
E^{(2)}[\Phi,\phi] &\= 2\pi\,\Tr\bigl\{
\phi^\+\,\Delta\phi\ -\ \phi^\+\,(\Phi\Delta\Phi^\+)\;\phi \} \\[4pt]
&\= 2\pi\,\Tr\bigl\{
\phi^\+\,\bigl(\begin{smallmatrix}1&0\\0&1\end{smallmatrix}\bigr)\,
\Delta\phi\ -\
\phi^\+\,\bigl(\begin{smallmatrix}0&0\\0&1\end{smallmatrix}\bigr)\,
2Q \bigl( |Q{-}1\>\<Q{-}1| + |Q\>\<Q| \bigr)\,\phi \bigr\} \ .
\end{aligned}
\end{equation}
Since the action of the Hessian is obviously diagonal, we find that
\begin{equation}
\phi \= \begin{pmatrix} \phi^{(1,1)} & \phi^{(1,2)} \\
                        \phi^{(2,1)} & \phi^{(2,2)} \end{pmatrix}
\qquad\Longrightarrow\qquad
E^{(2)}[\Phi,\phi] \= 
\sum_{i,j=1}^2 E^{(2)}\bigl[\Phi^{(i,i)},\phi^{(i,j)}\bigr] \ ,
\end{equation}
which reduces the fluctuation analysis to a collection of abelian cases.
For the case at hand, the Hessian in the $(1,.)$ sectors is given by
the Laplacian and thus non-negative, while in each $(2,.)$ sector 
it is identical to the Hessian for the abelian rank-$Q$ diagonal BPS case. 
Hence, the relevant results of the previous section carry over completely.

This mechanism extends to any diagonal U(2) background (not necessarily BPS),
mapping the U(2) fluctuation spectrum to a collection of fluctuation spectra
around corresponding diagonal U(1) configurations.
The fluctuation problem around non-diagonal backgrounds, in contrast, is not
easily reduced to an abelian one, except when the background is related to
a diagonal one by a rigid symmetry, as defined in (\ref{globalsym}), 
(\ref{globaltrans}) and (\ref{globalrot}). Only a small part of the moduli,
however, is generated by rigid symmetries, as our example of $U(\mu)$ in
(\ref{Umu}) demonstrates even for $Q{=}1$.
Finally, it is straightforward to extend these considerations to the 
general U($n$) case.

\section{Conclusions} 
\noindent
After developing a unified description of abelian and nonabelian 
multi-solitons in noncommutative Euclidean two-dimensional sigma models with
U($n$) or Grassmannian target space, we have analyzed the issue of their 
stability. Thanks to the BPS bound, multi-solitons in Grassmannian sigma models
are always stable. As in the commutative case, their imbedding into a unitary
sigma model renders them unstable however, as there always exists one negative
eigenvalue of the Hessian which triggers a decay to the vacuum configuration. 

Our results are concrete and complete for abelian and nonabelian $Q$-soliton 
configurations which are diagonal in the oscillator basis (or related to 
such by global symmetry).
For this case, we proved that the spectrum of the Hessian consists of the
essential spectrum $[0,\infty)$ and an eigenvalue $\lambda_Q$ of multiplicity
one with $-2Q<\lambda_Q<0$. 
This assertion was confirmed numerically, and the value of $\lambda_Q$ 
was given as a zero of a particular function composed of monomials in~$\lambda$
and the special function $\e^{-\lambda}\operatorname{li}(\e^{\lambda})$.

Furthermore, the complete set of zero modes of the Hessian was identified.
Each abelian diagonal $Q$-soliton background is characterized by a diagonal
projector~$P$ of rank~$Q$, whose image and kernel trigger a decomposition of  
the space of fluctuations into three invariant subspaces, namely
$u(\text{im}P)$, $u(\text{ker}P)$ and $\diff\,\text{Gr}(P)$. In addition, 
every side diagonal together with its transpose is seperately invariant 
under the action of the Hessian. This leads to a particular distribution 
of the admissible zero modes of the Hessian, displayed here for the example 
of~$Q{=}4$: 
\vskip-5mm
\begin{figure}[H]
\psfrag{phi=}{$\phi\ =\ $}
\psfrag{im P}{$\scriptstyle{\text{im}P}$}
\psfrag{ker P}{$\scriptstyle{\text{ker}P}$}
\psfrag{d Gr(P)}{$\scriptstyle{\text{d\,Gr($P$)}}$}
\psfrag{u(im P)}{$\scriptstyle{u(\text{im}P)}$}
\psfrag{u(ker P)}{$\scriptstyle{u(\text{ker}P)}$}
\begin{center}
\includegraphics[origin=ct,width=80mm]{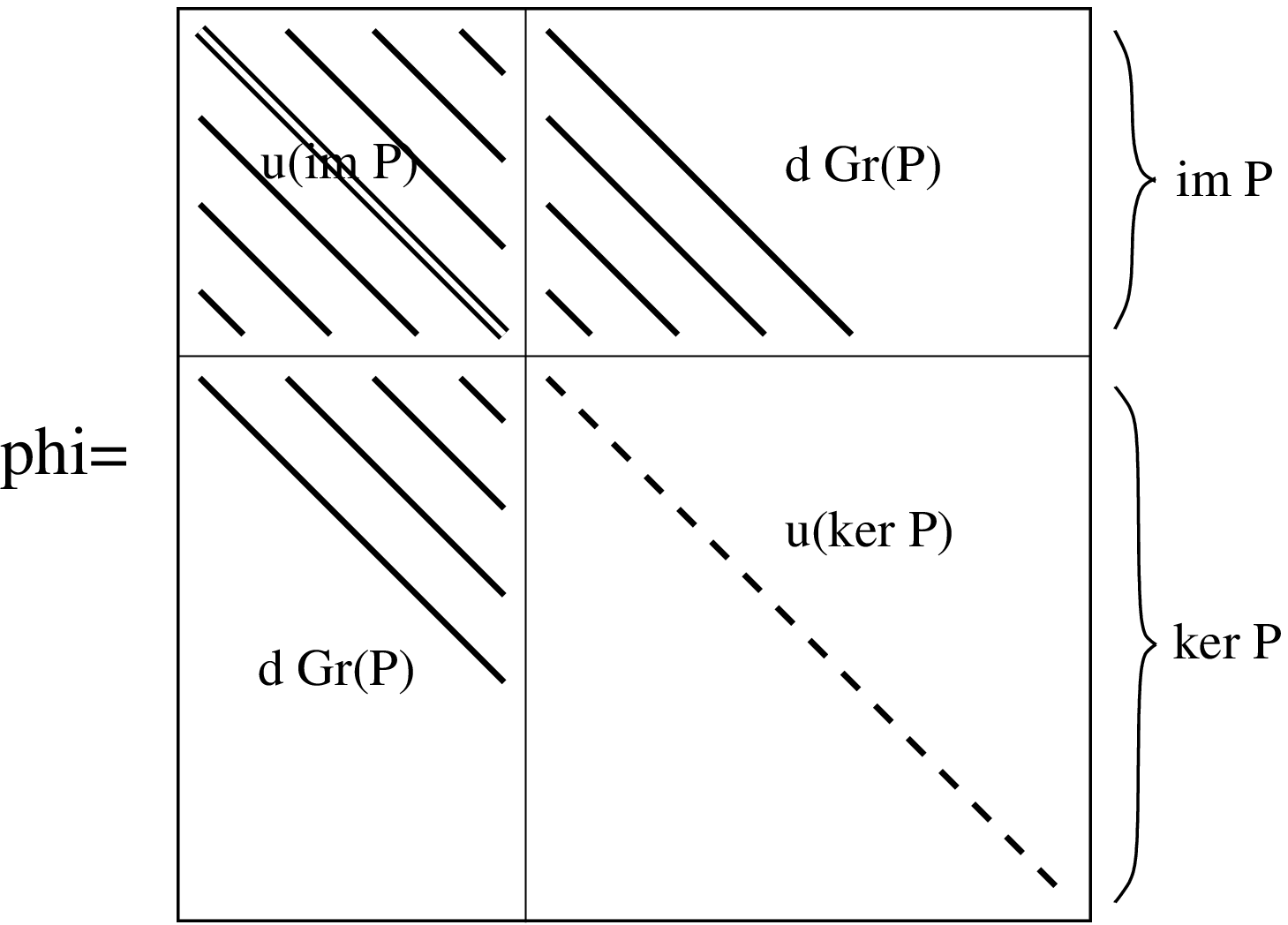}
\end{center} 
%\caption{}
\end{figure} 
\vskip-5mm \noindent
where the double line denotes the single negative eigenvector, 
each solid diagonal segment represents a real normalizable zero mode, 
the dashed line depicts an admissible non-normalizable zero mode, 
and empty areas do not contain admissible zero modes. 
In addition, each side diagonal features a non-admissible zero mode 
at the edge of the continuous part $[0,\infty)$ of the spectrum.
We plot the complete spectrum of the Hessian at~$Q{=}4$ 
(cut off at size $m_{\text{max}}{=}30$) for each invariant subspace
$\Ecal_k^{\text{Gr}(P)}$ (boxes), $\Ecal_k^{\text{im}P}$ (stars), 
$\Ecal_k^{\text{ker}P}$ (crosses) and $\Ecal_0$ (circles), up to $k{=}6$:
\vskip-5mm
\begin{figure}[H] 
\psfrag{$k$}{$\scriptstyle{k}$}
\psfrag{$lambda$}{$\scriptstyle{\lambda}$}
\begin{center}
\includegraphics[origin=ct,width=80mm]{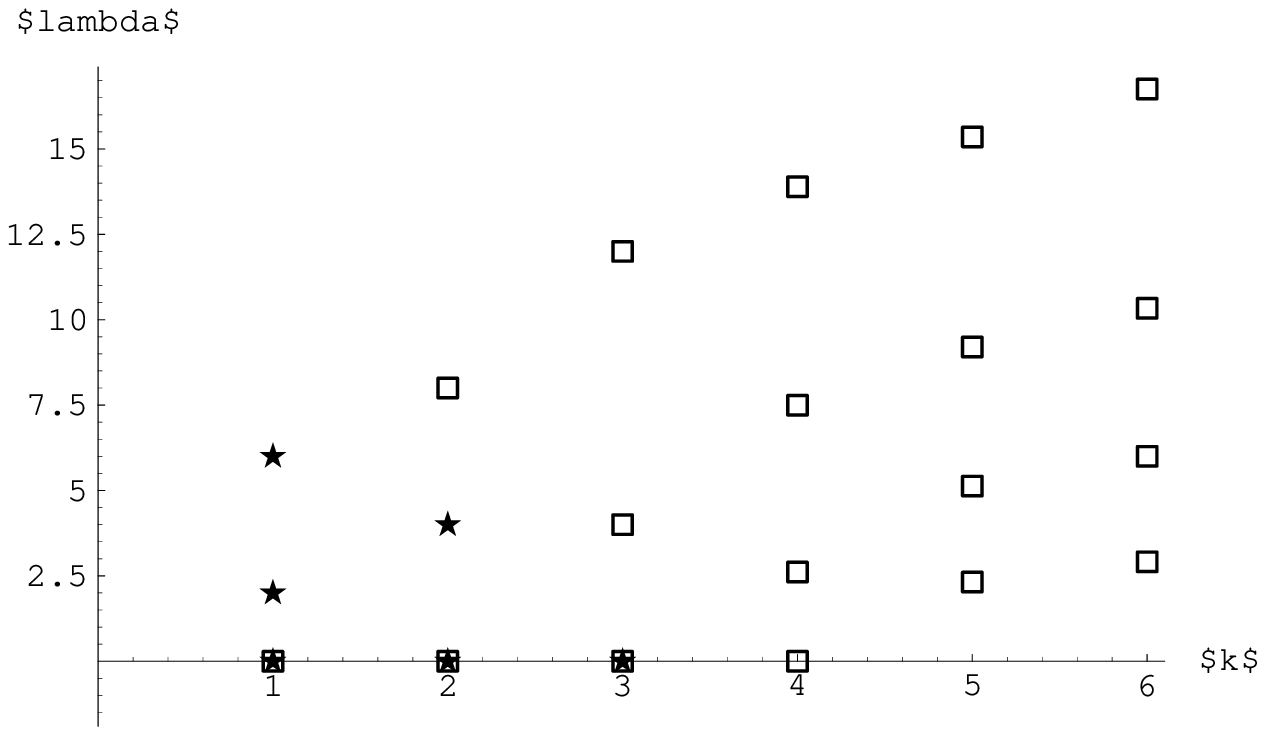}
\hspace{\fill}
\includegraphics[origin=ct,width=80mm]{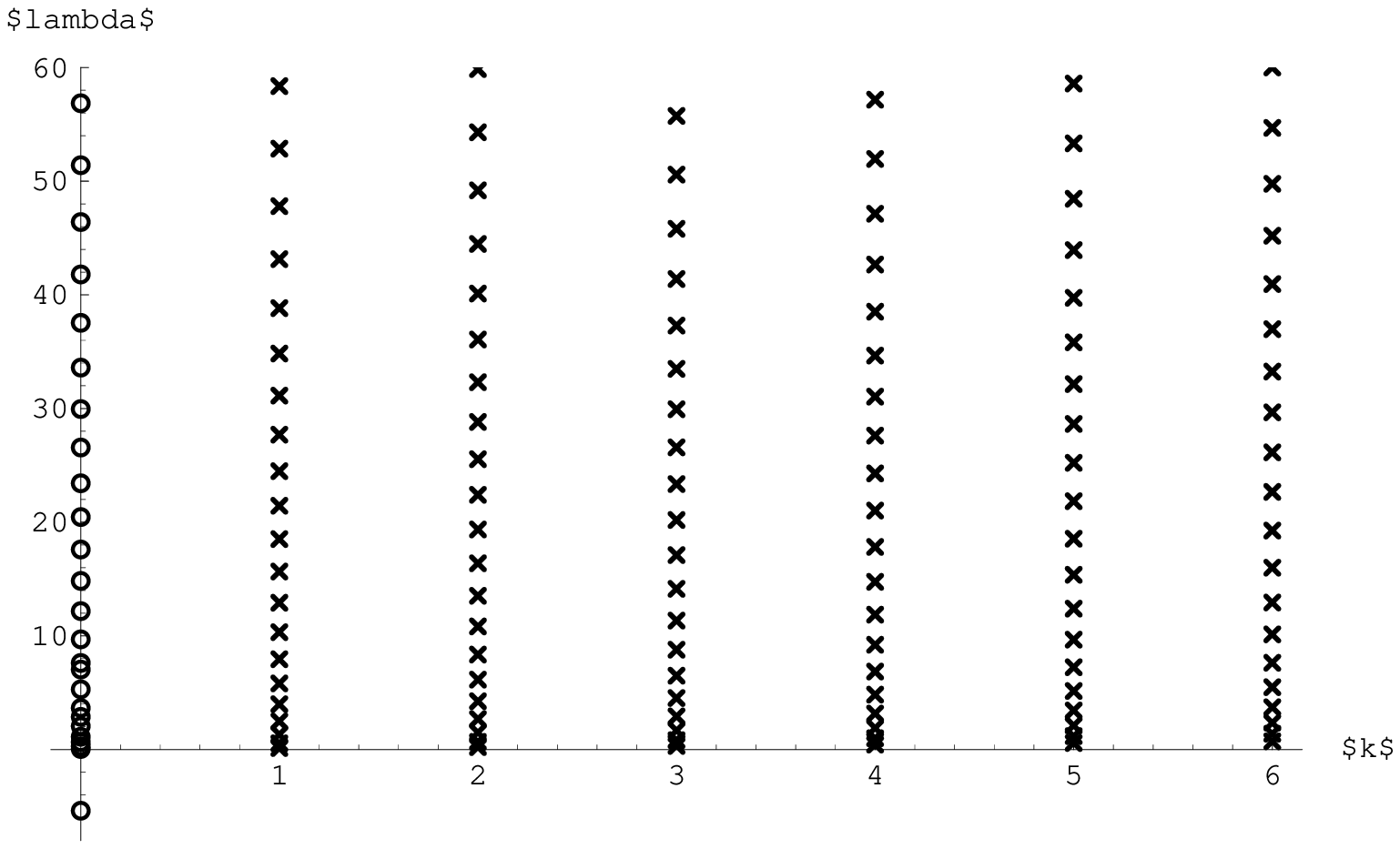}
\end{center}
\caption{Eigenvalues $\lambda$ of the cut-off Hessian (size~30) for $Q{=}4$
in subspaces $\Ecal_k$}
\end{figure}
Since most soliton moduli are not associated with global symmetries,
our explicit results for diagonal backgrounds do not obviously extend to 
generic (non-diagonal) backgrounds. We have not been able to compute the
fluctuation spectrum across the entire soliton moduli space. 
The only exception is the abelian single-soliton solution which always is 
a translation of the diagonal configuration and therefore fully covered 
by our analysis. Already for the case of two U(1) solitons, the unitary
transformation which changes their distance in the noncommutative plane
is only partially known.

This leads us to a number of unsolved problems. The most pressing one seems to
be the extension of our fluctuation analysis to non-diagonal backgrounds.
Next, following the even zero modes one can now find new non-BPS solutions.
Also, it is worthwhile to investigate the commutative limit of the Hessian
and its spectrum. Another interesting aspect is the existence of infinite-rank
abelian projectors, i.e.~via an infinite array of coherent states, associated
with BPS solutions of infinite energy. Furthermore, some technical questions
concerning the admissible set of fields and their fluctuations have remained.
Finally, it would be rewarding to extend the geometrical understanding of
sigma models to the noncommutative realm.

\bigskip

\noindent
{\large{\bf Acknowledgements}} \\
The authors are indebted to A.D.~Popov for numerous fruitful discussions
and thankful to R.~Wimmer and M.~Wolf for comments on the manuscript.
This work was partially supported by the Deutsche Forschungsgemeinschaft (DFG)
and the joint RFBR--DFG grant 436~RUS~113/669/0-2. In addition, 
A.V.D.~acknowledges support by the Russian Foundation for Basic Research, 
grants no.~04-01-00236 and 05-01-00981, as well as the Program in support of 
leading scientific schools no.~NSh-2040.2003.1. He is also very grateful to the
Institut f\"ur Theoretische Physik at Universit\"at Hannover for hospitality.

%\pagebreak

\end{document}